\documentclass[fleqn,10pt]{wlscirep}

\usepackage{graphicx}
\usepackage{dcolumn}
\usepackage{bm}
\usepackage{amsmath}
\usepackage{amssymb}
\usepackage{color} 
\usepackage{pifont} 
\usepackage{stmaryrd} 
\usepackage{wasysym} 
\usepackage{float}
\usepackage{pdfsync}

\title{Griffiths phase and long-range correlations in a biologically motivated visual cortex model}

\author[1,+]{M.~Girardi-Schappo}
\author[1]{G.~S.~Bortolotto}
\author[1]{J.~J.~Gonsalves}
\author[2]{L.~T.~Pinto}
\author[1,*]{M.~H.~R.~Tragtenberg}
\affil[1]{Departamento de F{\'i}sica, Universidade Federal de Santa Catarina, 88040-900, Florian{\'o}polis, Santa Catarina, Brazil}
\affil[2]{Departamento de Engenharia Qu{\'i}mica e Engenharia de Alimentos, Universidade Federal de Santa Catarina, 88040-900, Florian{\'o}polis, Santa Catarina, Brazil}

\affil[+]{girardi.s@gmail.com}
\affil[*]{marcelotragtenberg@gmail.com}
\affil[$ $]{doi:10.1038/srep29561}


\begin{abstract}
Activity in the brain propagates as waves of firing neurons,
namely avalanches. These waves' size and duration distributions
have been experimentally shown to display a stable power-law profile,
long-range correlations and $1/f^{b}$ power spectrum
\textit{in vivo} and \textit{in vitro}.
We study an avalanching biologically motivated model of mammals visual cortex and 
find an extended critical-like region -- a Griffiths phase -- characterized by divergent susceptibility and
zero order parameter. This phase lies close to the expected experimental value of the \textit{excitatory postsynaptic potential}
in the cortex suggesting that critical behavior may be found in the visual system.
Avalanches are not perfectly power-law distributed, but it
is possible to collapse the distributions and define a cutoff avalanche size
that diverges as the network size is increased inside the critical region.
The avalanches present long-range correlations and $1/f^{b}$ power spectrum, matching experiments.
The phase transition is analytically determined by a mean-field approximation.
\end{abstract}

\def\camII@III{II/III}
\def\camIV{IVC$\beta$}
\def\camVI{VI}
\def\camLGN{LGN}
\def\mV{\ensuremath{~\textnormal{mV}}}

\def\virgula{\ \textnormal{,}} 

\definecolor{red1}{RGB}{249,42,42}
\definecolor{red2}{RGB}{194,29,29}
\definecolor{red3}{RGB}{146,18,18}
\definecolor{red4}{RGB}{89,5,5}
\definecolor{blue1}{RGB}{42,42,249}
\definecolor{blue2}{RGB}{29,29,194}
\definecolor{blue3}{RGB}{18,18,146}
\definecolor{blue4}{RGB}{5,5,89}
\definecolor{gray1}{rgb}{0.9,0.9,0.9}
\definecolor{gray2}{rgb}{0.8,0.8,0.8}
\definecolor{redC}{RGB}{210,45,45}
\definecolor{blueC}{RGB}{112,179,250}
\definecolor{greenC}{RGB}{99,199,113}
\definecolor{purpleC}{RGB}{194,129,201}

\begin{document}

\flushbottom
\maketitle
%
%
\thispagestyle{empty}

\section*{Introduction}

The most important feature of a living complex system to survive is adaptability. In order to adapt, the organism cannot
be inflexible, but also cannot act randomly. The border of chaos or a critical behavior seems to be the best evolutionary
choice to survival beings~\cite{chialvoArx2008,hidalgoEvolSOC2014}.
Turing conjectured a similar idea referring to 
our minds and learning machines~\cite{turing1950}. 

More recently, brain criticality has become a trendy research
subject~\cite{chialvoReview,plenzBenefits,beggsEditorial}. Some believe that the brain became
critical by self-organization and selection~\cite{chialvoReview}, analogously to the
sandpile self-organized criticality (SOC)~\cite{bakPRL}. 
Criticality has the advantages of maximizing the response
dynamic range of neural networks~\cite{kinouchiCopelli},
optimizing memory and learning processes~\cite{socPlasticity},
the computational power of the brain~\cite{beggsCritical}
and information processing flexibility~\cite{mosqueiro2013}.

The brain critical state is usually experimentally characterized via power-law (PL) distributed neuronal
avalanches~\cite{beggsPlenz2003,ribeiroCopelli,plenzAvalV12010,priesemannSub2014,shewV1Aval2015}
or diverging long-range spatial or temporal correlations~\cite{linkenkaerSOC,haimoviciRestState2013}.
On the other hand, criticality in (non-)equilibrium Statistical Mechanics is only rigorously defined 
via the PL convergence to zero
of an order parameter and the simultaneous PL divergence of its associated
susceptibility~\cite{barberFSS1983,stanley,noneqPhaseTrans2008}.
In addition, finite systems avalanche size distributions present cutoffs, which also must diverge 
according to PLs on the critical state~\cite{pruessnerSOC2012}.
These PL critical exponents must follow well-defined scaling
relations~\cite{stanley,munozExpAbsState1999,odorReview2004,noneqPhaseTrans2008}.
Notice, however, that some authors do not take into account the finite-size scaling (FSS) of avalanche
distributions cutoff~\cite{shewV1Aval2015}.
Also, only a few authors use an order parameter--susceptibility pair to probe for
criticality in Neuroscience~\cite{plenzExp2013,munozGrif2013}.

Here we show that there is a Griffiths~\cite{griffiths1969,vojtaGrifRev2006,munozGrif2013}
phase (GP) in the non-equilibrium percolation-like phase transition of the visual processing activity
as a function of the \textit{excitatory postsynaptic potential} (EPSP). The GP is known to emerge
from rare over-active region effects due to the quenched disorder of the network. One of the macroscopic consequences
of such effects is the appearance of an extended region of critical behavior in the phase diagram of the system.
This result for the visual system is consistent with recent findings that suggest that excitable systems running over
the connectome structure (a whole brain network) present GP~\cite{munozGrif2013}.
Our control parameter is the EPSP and the GP is found close to the experimental values of EPSP in the
cortex~\cite{williamsV12002,songEPSP2005,lefortEPSP2009}. The EPSP threshold for a complete network activation is determined
via a mean-field approximation and is close to the expected numerical and experimental values of this parameter.

We define the density of activated neurons as an order parameter
and verify that it converges to zero whereas its associated susceptibility diverges inside the while GP by
applying standard FSS technique. This scaling rigorously defines the critical phase transition in our model.
Our order parameter is usual for absorbing state phase transitions~\cite{janssenEPDynP2004,noneqPhaseTrans2008}.
Additionally, we show that the avalanches are PL distributed with a cutoff that scales with the system size inside the GP.
We also study the visual system avalanches correlations and power spectrum.
Throughout the critical region, the power spectrum of avalanche time series
has the form $1/f^{b}$, with $0.2\leq b\leq1.3$
as experimentally expected~\cite{novikov1fB1997,teichcat97,linkenkaerSOC,henrie1fV12005,hermes1fV12014}.
The activity time series inside the critical region also presents long-range correlations yielding
\textit{Detrended Fluctuation Analysis}~\cite{stanleyDFA1995,linkenkaerSOC,linkenkaerDFA2012}
(DFA) exponent $g\gtrsim0.7$.

Avalanche distributions presenting PL alone have been questioned as insufficient
evidence for identifying the critical regime~\cite{beggsCritical,hartleyISuscep2013,destexhePL2015}
since there may be critical dynamical systems which have no PL distributed
avalanches~\cite{hartleyISuscep2013} and non-critical dynamical systems that present PL distributed
avalanches~\cite{destexhePL2015}.
Our model's avalanche distributions have PL shape inside the GP and even outside it.
Thus, we show another example that PL distributed avalanches is not
a sufficient condition for criticality. More than that, we use an order parameter--susceptibility
pair as usually done for phase transitions to probe for
criticality~\cite{stanley,munozExpAbsState1999,odorReview2004,noneqPhaseTrans2008}.

We chose to model the visual cortex because it has a well-known anatomy and
function~\cite{okuskyV11982,lundV11984,callawayV1Review1998,yabutaV11998,albrightNeuroRev2000}.
The visual cortex also has a valuable data set available that we may use to benchmark our model results,
such as the power spectrum~\cite{novikov1fB1997,teichcat97,linkenkaerSOC,henrie1fV12005,hermes1fV12014}
and a few avalanche experiments~\cite{ribeiroCopelli,plenzAvalV12010,shewV1Aval2015}.
We focus on understanding the dynamics of the signal propagation and the avalanche activity
related to the network disordered structure.
Our model is biologically motivated in the sense that the network we study here is layered, columnar, and
recurrent, resembling the architecture of the visual cortex~\cite{albrightNeuroRev2000}
(full details of the network are given in the Supplementary Material).
Also, our parameters are either fitted to experiments (such as the attenuation
constant~\cite{williamsV12002}) or
have been experimentally measured
(such as the structural parameters\cite{okuskyV11982,lundV11984,callawayV1Review1998,yabutaV11998}).

The avalanches spontaneously emerge after a flash stimulus presented to the 
the model's retina, instead of artificially imposing an abstract Poisson 
stimulus as in most of the brain critical models
(see~\cite{kinouchiCopelli,socPlasticity,levina,ariadneDynSyn2015,ribeiroCopelli,girardiAva,beggsSOqC2014,shewV1Aval2015}
to cite a few). Avalanche dynamics is thus essential for the reliable signal propagation in our model.

We describe key features of the considered model in the following section. In the introduction of the Results
section, we define the Griffiths phase and briefly discuss its origin in the model,
the tools we use to characterize it (order parameter and susceptibility)
and the observables of interest (avalanches, processing time, avalanche distributions and
correlation measurements). Results are presented and discussed in the subsections of Results
section. We finish the paper by briefly reviewing our main results and pointing how
they relate to experiments and to SOC models in the Concluding Remarks section.

\section*{Model}

\begin{figure*}[t!]
	\includegraphics[width=160mm]{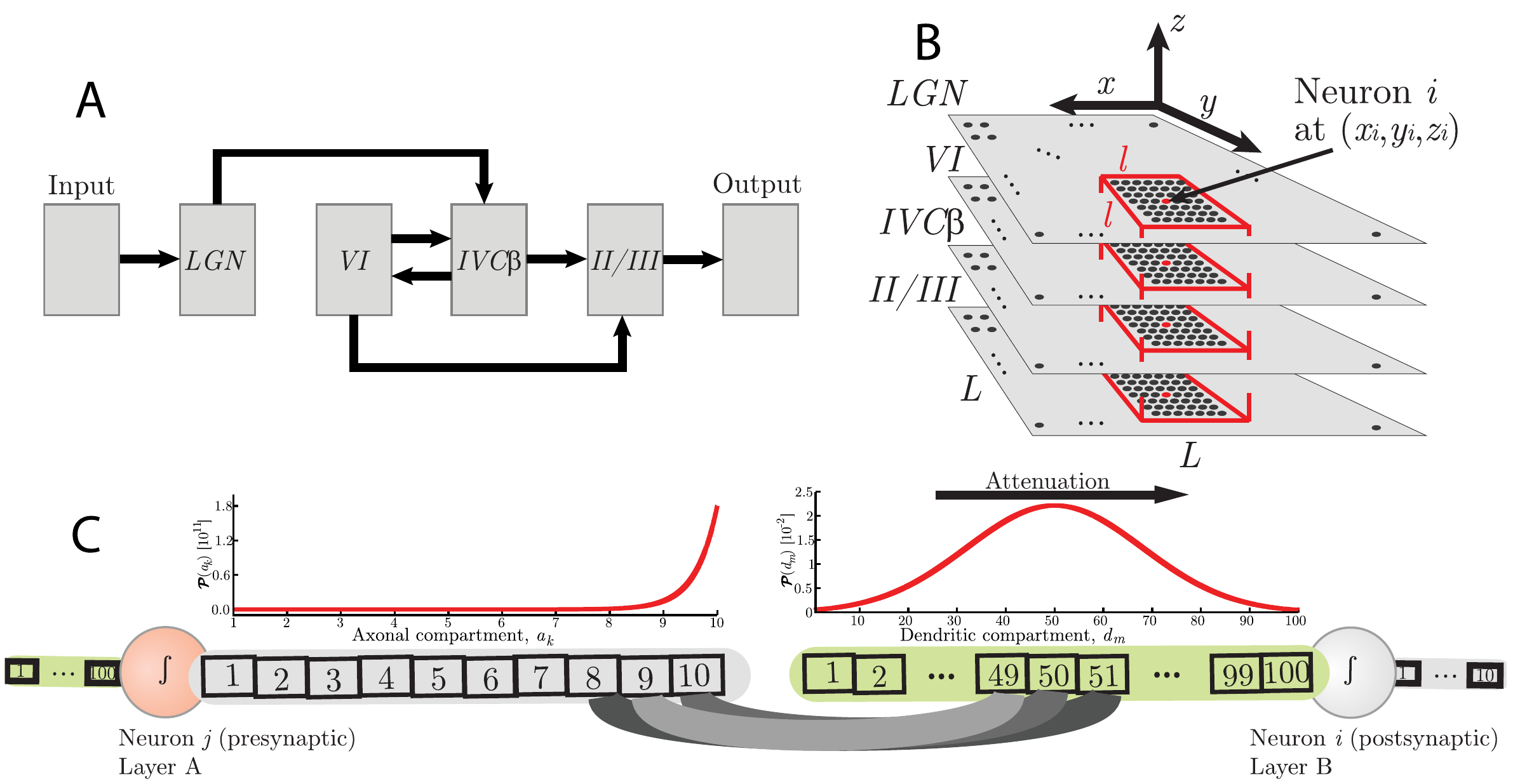}
	\caption{\label{fig:modeloV1}Elements of the V1 model.
	A: Architecture of the network.
    B: Spatial organization of the network of $N=4L^2$ neurons. The columnar structure is highlighted in red. There is a column of size $N_c=4l^2=196$ neurons centered on each neuron of the network.
	C: Compartmental scheme of neurons.
	The probability, $\mathcal{P}(a_k)$, of choosing a presynaptic axonal compartment, $a_k$, of any neuron is exponential such that most of the synapses start from the end of the axon (left). The probability, $\mathcal{P}(d_m)$, of choosing a dendritic postsynaptic compartment, $d_m$, is Gaussian with mean $50$ and standard
    deviation $10$, so that most of the synapses lay in the middle of the dendrite (right).}
\end{figure*}

This model was originally developed by Andreazza~\& Pinto~\cite{andreazzaSim2006}
in order to study the signal propagation dynamics in the visual cortex of mammals.
They detected some sort of phase transition which will be analyzed in details throughout this work.
The model is composed of six interconnected square layers.
The signal propagates directionally from the retina (the Input layer) to the
secondary visual cortex (Output layer). The other four layers have lateral size $L$ and
are selected from the form recognition pathway:
the lateral geniculate nucleus (\camLGN, from the thalamus) and the layers \camII@III, \camIV\ and
\camVI\ from the primary visual cortex (V1). The
architecture of the network is illustrated in Figure~\ref{fig:modeloV1}A, where arrows point the direction
of the connections through which the signal propagates. The direction of the connections characterizes
the adjacency of layers.
The \camLGN\ layer consists of only its parvocellular neurons. Their
synapses are mostly connected to V1 layer
\camIV~\cite{okuskyV11982,lundV11984,callawayV1Review1998,yabutaV11998}.
The synaptic buttons density over the dendrites is also based on
experiments~\cite{okuskyV11982,lundV11984,callawayV1Review1998,yabutaV11998}. 
The bulk of the network (layers \camLGN, \camVI, \camIV\ and \camII@III\ together) has $N=4L^2$
neurons.
The Input layer (composed of photoreceptors) and the Output layer (composed of axon terminals
that connect to secondary visual cortex) have $N_{io}=(10L)^{2}$ elements each.
Each neuron $i$ of the four internal layers is composed of a dendrite with 100 compartments, the soma, and
an axon with 10 compartments.


The network is built in four steps: (a) for each neuron of the network, a postsynaptic neuron is chosen
from an adjacent layer using a two-dimensional Gaussian distribution inside a limited excitatory field of size $l^2$;
(b) an axonal compartment is chosen from the presynaptic neuron using an exponential probability distribution
plotted in Figure \ref{fig:modeloV1}C (left); (c) a dendritic compartment of the postsynaptic neuron is
chosen using a Gaussian distribution centered in the middle of the dendrite, as in Figure \ref{fig:modeloV1}C (right);
and (d) a synapse is formed by connecting the chosen axonal compartment and the chosen dendritic compartment.
There is a different number of outward synapses per presynaptic neuron depending on the presynaptic layer
(see tables in Supplementary Material).
The randomly chosen postsynaptic neurons and the randomly chosen pairs
of presynaptic axonal compartments and postsynaptic dendritic compartments give rise
to quenched disorder.

This structure generates a directed path of square columns of highly connected neurons
for the signal propagation centered in each neuron of each layer.
Each column has $N_c=4l^2\approx200$ neurons (see Figure~\ref{fig:modeloV1}B).
As an example, a network with $L=99$ (the largest considered size in this work) has approximately
$N=4\times10^4$ neurons and $32.5\times10^6$ synapses in total. It is thus computationally
costly to simulate larger networks. Once the network structure is built, it is kept fixed for a single signal
propagation dynamics.

The EPSP (in the dendrites) and the action potential (in the axons) advance one compartment per time step $t$,
coming from the dendrites through the soma to the last axonal compartment.
The variables $d_m^{(i)}(t)$ [Equation \eqref{eq:compDendritoV1}], $v_i(t)$
[Equation \eqref{eq:compSomaV1}] and $a_k^{(i)}(t)$ [Equation~\eqref{eq:compAxonioV1}] represent the
local values of the EPSP sum (for each dendritic compartment $m$)
or the action potential (for each axon compartment $k$)
in the membrane of neuron $i$ at time $t$, respectively.

These rules may be summarized in the following equations:
\begin{equation}
\label{eq:compDendritoV1}
\begin{array}{ll}
    d_1^{(i)}(t+1)&=\lambda E\sum\limits_{j,n}{a_n^{(j)}(t)}\virgula\\
    d_k^{(i)}(t+1)&=\lambda\left[ d_{k-1}^{(i)}(t)+E\sum\limits_{j,n}{a_n^{(j)}(t)}\right],\textnormal{for }k>1,
\end{array}
\end{equation}
\begin{equation}
\label{eq:compSomaV1}
v_i(t+1)=\left\{
\begin{array}{ll}
    \Theta\left(d_{100}^{(i)}(t)-v_T\right) & \textnormal{, if }v_i(t)=0\virgula\\
    -R & \textnormal{, if }v_i(t)=1\virgula\\
    v_i(t)+1 & \textnormal{, if }v_i(t)<0\virgula
\end{array}
\right.
\end{equation}
\begin{equation}
\label{eq:compAxonioV1}
\begin{array}{ll}
    a_1^{(i)}(t+1)&=\Theta\left(v_i(t)\right)\virgula \\
    a_k^{(i)}(t+1)&=a_{k-1}^{(i)}(t), \textnormal{for }k>1,
\end{array}
\end{equation}
where $E>0$ is the EPSP,
$\lambda=0.996$ is the dendritic attenuation constant
(chosen to match experimental attenuation~\cite{williamsV12002}),
$\Theta(x)$ is the Heaviside step function and
$v_T=10\mV$ is the firing threshold needed to induce an action potential~\cite{kandel2012}.
After the soma spikes, it is reset to the value
$R$ which represents the refractory period (in time step unit). $R$ is set to avoid
self-sustained activity in the interlayer loops (between layers \camIV\ and \camVI).
The signal propagates in the direction of increasing $k$ (as $t$ increases).
The control parameter $E$ sets the level of excitation,
whereas $\lambda$ sets the level of dissipation.
The double sum in Equation~\eqref{eq:compDendritoV1} is over each of the axonal compartments $n$ of the
presynaptic neuron $j$ connected to the dendritic compartment $k$ of the postsynaptic neuron $i$.
 
Every neuron initial condition is given by
$a_k^{(i)}(0)=d_m^{(i)}(0)=v_i(0)=0\ \forall\ (k,m,i)$.
The time scale of the model is arbitrary, so
the spike signal of the soma takes about~1~ms. The different number
of compartments for dendrites and axons copes with their different velocities for
the signal propagation~\cite{deutschNervous1993}.

\section*{Results and discussion}

For each $E$ and $L$, the activity is initiated by flashing once a square of $30\times30$ photoreceptors in the center of
the Input layer (which, in turn, will activate a region of $3\times3$ neurons in the \camLGN).
We also consider different positions for the initial stimulus, such as flashing a square near
the corner of the Input layer in order to verify if results are stable.
Notice that this procedure is similar to that used in the study of epidemic spreading
in systems having absorbing phase transitions, in which a single site (or a small fraction of sites)
is initially infected~\cite{noneqPhaseTrans2008}. However, in epidemic spreading models
each site is a simple three or four states cellular automaton and the network is generally regular
and hypercubic~\cite{noneqPhaseTrans2008}.

\begin{figure*}[t!]
	\includegraphics[width=170mm]{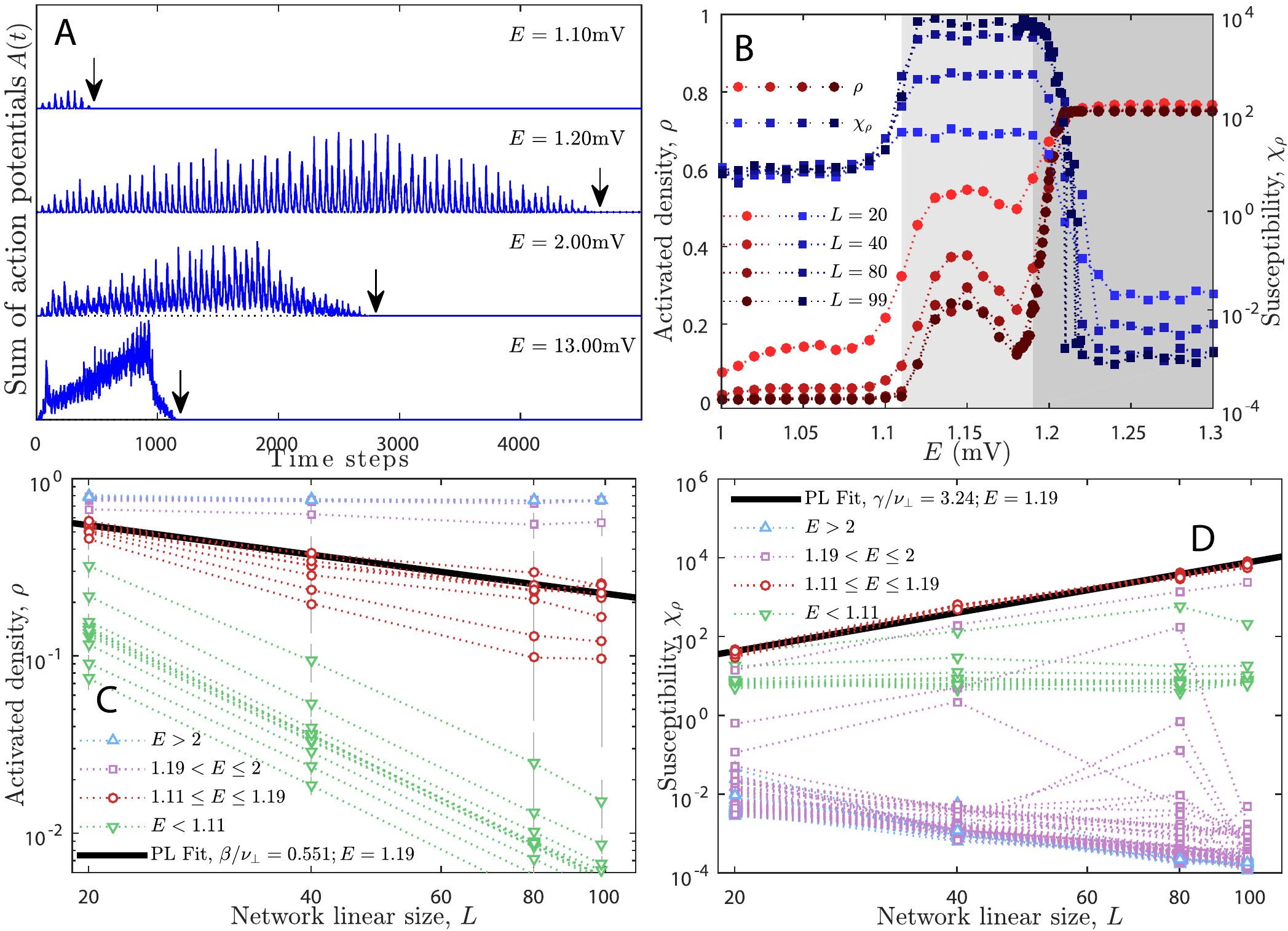}
	\caption[Network activity, order parameter and susceptibility.]{\label{fig:tsOrderSusc}Network activity,
    order parameter and susceptibility.
	Panel~A: Temporal profile of the avalanches
	for many EPSP. Arrows mark the processing time $T$ of the network.
    For $E=1.1\mV$, only very small avalanches occur; $E=1.2\mV$ shows many small avalanches;
    for $E=2.0\mV$ there is a dominating avalanche (compare to the dotted line that marks zero activity);
    and for $E=13.0\mV$ the dominating avalanche takes over all the dynamics.
	Panel~B: Density of activated neurons $\rho$ during total activity time and its associated susceptibility
		$\chi_{\rho}=N(\langle\rho^2\rangle-\langle\rho\rangle^2)/\langle\rho\rangle$~\cite{ferreiraSus2012}
		as function of $E$ for many $L$. Red circles
        (\textcolor{red1}{\ding{108}},\textcolor{red2}{\ding{108}},\textcolor{red3}{\ding{108}},\textcolor{red4}{\ding{108}})
        indicate $\rho$ and blue squares
        (\textcolor{blue1}{\ding{110}},\textcolor{blue2}{\ding{110}},\textcolor{blue3}{\ding{110}},\textcolor{blue4}{\ding{110}})
        indicate $\chi_{\rho}$; the larger $L$ the darker the color shade.
        White background indicate the inactive phase, light gray background indicates the critical (Griffiths) phase with
        diverging $\chi_{\rho}$ and dark gray background indicates the percolating phase.
	Panels~C,D: FSS of $\rho$ order parameter~\cite{vojtaGrifRev2006,munozGrif2013} yielding
	scaling exponent $\beta/\nu_{\bot}=0.55(3)$ -- Equation~\eqref{eq:activedens} -- and
	FSS of $\chi_{\rho}$ yielding
	scaling exponent $\gamma/\nu_{\bot}=3.1(1)$ -- Equation~\eqref{eq:suscept}; fits performed on the transition
    point $E_c=1.19\mV$; (\textcolor{greenC}{$\triangledown$}) inactive phase,
    (\textcolor{redC}{\Circle}) critical phase,
    (\textcolor{purpleC}{$\Box$}) and (\textcolor{blueC}{$\vartriangle$}) percolating phases.
	Vertical bars are standard deviation and dotted lines are only guides to the eyes.}
\end{figure*}

In the present case, the extended body of
each neuron (or site) and the disordered structure of the network generates a spreading
pattern that works like a branching process~\cite{girardiV1conf2015}.
Nevertheless, the adjusted $R$ ensures that the activity will always die out
and our model (as considered in this paper) has no active stationary state (see in Figure~\ref{fig:tsOrderSusc}A that activity
eventually fades away for every $E$).
After the activity of a single stimulation has ended, the amount of neurons that fired is $N_R$.
The configuration of the network made of $N_R$ activated neurons and $N-N_R$ inactivated neurons is an absorbing state.
As we vary $E$, we notice that a larger fraction of neurons gets activated by the
initial stimulus. For $E>E_{th}$, the activity will definitely percolate the network
before getting extinguished (i.e. $N_R=N$). Our model, thus, presents an absorbing state phase
transition from a non-percolating inactive phase (small $E$) to a dynamically percolating phase (large $E$)~\cite{grassbergerGEP1983}.
We employ a mean-field analysis to determine $E_{th}$ in the next subsection. Intermediary values of $E$ yield vanishing
non-zero probability of percolation.

After the activity is ceased, another trial is started by (a) rewiring the network
according to the rules
described in last section and in the Supplementary Material; (b) resetting the state of
the neurons; and (c) flashing the same square stimulus to the network.
This procedure is repeated many times for each $E$ and $L$.
Therefore, the averages and variances presented in this section are calculated over these
several trials of the network quenched disorder in the connections between neurons.
The disorder of the system creates rare region fluctuation
effects on the size $N_R$ of the percolation cluster for a finite range of the $E$
parameter. The system then presents a GP in this interval~\cite{griffiths1969,vojtaGrifRev2006,munozGrif2013}.
Figure~\ref{fig:tsOrderSusc}A shows the
temporal profile of activity $A(t)$ (the sum of every soma spike in the network for a given
time step) for a single trial of the network with $L=99$ and four different $E$.

The temporal profile of activity comprises interspersing small and large peaks of activity
that spontaneously emerge (see Figure~\ref{fig:tsOrderSusc}A)
due to the delay caused by the propagation of potentials through dendrites and axons.
Thus, the separation of the peaks (i.e. the separation of \textit{time scales}) in our model output
data is not externally imposed, as commonly done in SOC models~\cite{pruessnerSOC2012}.
This emergent separation of activity intervals provides a natural way to define the avalanches of our model.
The observed avalanches are correlated because the peaks follow from each other in
an organized temporal sequence.
This is different from the Poisson independently generated
avalanches obtained in most models of absorbing state phase
transitions~\cite{kinouchiCopelli,socPlasticity,levina,ariadneDynSyn2015,ribeiroCopelli,pruessnerSOC2012,girardiAva,beggsSOqC2014}.

In such models, each avalanche is generated by a single stimulus and one avalanche is defined
as all the activity between two inactive absorbing states of the
system~\cite{munozExpAbsState1999,pruessnerSOC2012}. On the contrary, the dynamics of our model
allows us to define avalanches using the experimental protocol: here,
one avalanche is all the activity between two moments of silence in the network,
even if the system did not reach its absorbing state during such silent period.
In fact, when analyzing experimental data there is no guarantee that the system is in its inactive
absorbing state between consecutive avalanches because background activity of the
network is always neglected by applying \textit{ad hoc} thresholds to the
electrodes' signals~\cite{beggsPlenz2003,ribeiroCopelli,plenzAvalV12010,shewV1Aval2015}.
The size $s$ of the avalanche is the sum of all the activity $A(t)$ between two consecutive instants
in which $A(t)=0$. In fact, each $s$ is the area under each of the peaks in
Figure~\ref{fig:tsOrderSusc}A.
The distribution of avalanche sizes $s$ of our model will be discussed in the sections to follow.






Absorbing state phase transitions are commonly studied through the definition of
two order parameters~\cite{noneqPhaseTrans2008}: the density of active sites in the
active stationary state (which defines the critical exponent $\beta$); or
the percolation probability (which defines the critical exponent $\beta'$).
The second one is most commonly used for systems without active
stationary state. However, the density of activated sites by an initial stimulus
(i.e. the density of sites that pertain to the percolating cluster,
also known as debris density) is also used to describe phase transitions
without active absorbing states~\cite{janssenEPDynP2004}. Thus, we chose
to study our systems' phase transition using the density of activated
neurons, $\rho\equiv\langle N_R/N\rangle$, as our order parameter, because it may be
directly measured for each quenched disorder configuration of the network.
Notice that the average is taken over the many trials of the network for fixed $E$ and $L$:
for a given trial, a quantity $N_R$ of neurons has been activated after the activity dies out;
$\rho$ is then the average of $N_R/N$.
The amount of activated neurons for each trial is simply calculated by $N_R=\sum_{t=0}^{T}A(t)$,
where $T$ is the total propagation time (time from the moment the
activity is sparked until the moment it dies out). The quantity $T$ is also known as
mean survival time in absorbing phase transitions~\cite{noneqPhaseTrans2008}.

The variance of $\rho$ is regarded as a susceptibility,
$\chi\equiv N(\langle\rho^2\rangle-\langle\rho\rangle^2)$, and defines the critical
exponent $\gamma'$ for models presenting typical absorbing phase
transitions~\cite{noneqPhaseTrans2008,ferreiraSus2012}. Nevertheless, notice
that in such systems the variance of $\rho$ is often taken over its temporal fluctuations
in the absorbing active state~\cite{noneqPhaseTrans2008}. Our model does not have an active absorbing state, so the fluctuations
of $\rho$ arise from different trials of the network disorder for fixed $E$ and $L$, similarly to how $\rho$ is measured:
after each trial, a total density $N_R/N$ of activated neurons is left by the propagated network activity.
Thus, the variance is calculated over all these trials for each $E$ and $L$.
Since our network is not regular, we chose to study a modified version of the
susceptibility for complex networks,
$\chi_{\rho}=\chi/\langle\rho\rangle$, which defines a critical exponent $\gamma=\gamma'+\beta$~\cite{ferreiraSus2012}.
Notice that this $\gamma$ is not related to the average avalanche size of absorbing phase transitions.
The plot of the standard susceptibility $\chi$ is shown in the Supplementary Material. It presents the diverging pattern
with exponent $\gamma'$ expected for absorbing critical systems. However, the modified
susceptibility shows in a more neat way the extent of the GP.

Near the critical point, we may write the following
scaling functions~\cite{barberFSS1983,noneqPhaseTrans2008,vojtaGrifRev2006,munozGrif2013}:
\begin{align}
\label{eq:activedens}
\rho(E;L)&\sim L^{-\beta/\nu_{\bot}}\bar{\rho}(|E-E_c|L^{1/\nu_{\bot}})\ ,\\
\label{eq:suscept}
\chi_{\rho}(E;L)&\sim L^{\gamma/\nu_{\bot}}\bar{\chi}_{\rho}(|E-E_c|L^{1/\nu_{\bot}})\ ,
\end{align}
where $\beta$, $\nu_{\bot}$, and $\gamma=\gamma'+\beta$ are scaling exponents and 
$\bar{\rho}(x)$ and $\bar{\chi}_{\rho}(x)$ are universal scaling functions.
If equations~\eqref{eq:activedens} and~\eqref{eq:suscept} hold for the computational data, 
then a critical phase transition occurs on
$E_c$~\cite{barberFSS1983,stanley,odorReview2004,noneqPhaseTrans2008}.
The value $E_c$ marks the point in which the probability of appearing a percolation cluster
due to the system dynamics continuously changes from zero to positive~\cite{noneqPhaseTrans2008}.
This is the strong criterion to define criticality, although it is not commonly applied in neural networks models.
Only one work has experimentally applied it so far in the brain context~\cite{plenzExp2013}.
Notice that the exponent $\gamma'$ is calculated for the variance of $\rho$ over
the trials of the network, whereas the usual $\gamma'$ is obtained through
temporal fluctuations. Thus, the $\gamma'$ we calculate in this work might not
correspond to the usual $\gamma'$ of absorbing phase transitions.

We also compute the avalanche size distributions, $\mathcal{P}(s)\sim s^{-\alpha}$,
the complementary cumulative distributions, $\mathcal{F}(s)$,
the avalanches' autocorrelation, $C(t')$,
and power spectrum, $S(f)$,
the activity time series DFA, $F(\Delta t)$,
and measure the network processing time, $T(E;L)$. In the critical point, these quantities
may be written as:
\begin{align}
\label{eq:scaling}
\mathcal{F}(s)&\sim s^{1-\alpha}\ ,\\
\label{eq:corrFunc}
C(t')&\sim t'^{-\theta}\exp\left(-t'/\tau\right)\ ,\\
\label{eq:powerSpec}
S(f)&\sim f^{-b}\ ,\\
\label{eq:dfa}
F(\Delta t)&\sim(\Delta t)^g\ ,\\
\label{eq:proctime}
T\left(E;L\right)&\sim L^{\mu}\bar{T}\left(E\right)
\end{align}
where
$\alpha$ is the exponent of avalanche sizes distribution,
$t'$ is the time lag between two avalanches,
$\tau$ is the characteristic time of the autocorrelation exponential cutoff,
$\theta$ and $b$ are the autocorrelation and power spectrum
exponents of the \textit{avalanche time series}, respectively,
$g$ is the DFA exponent of the \textit{activity time series},
$\mu$ is a scaling exponent and
$\bar{T}(E)$ is a universal scaling function.
Since $S(f)$ is the Fourier transform of $C(t')$, $\theta+b=1$ for $t'\ll\tau$.

The cumulative avalanche size distribution provides a clearer and direct way to calculate the cutoff
$Z$ of the avalanche size distribution. We assume that
$\mathcal{P}(s)\sim s^{-\alpha}$ is valid for $s\leq Z$, therefore
\begin{equation}
\label{eq:cDistFit}
\mathcal{F}(s)\equiv\int_s^Z{\mathcal{P}(s')ds'}=c_1+c_2s^{-\alpha+1}\ ,
\end{equation}
and thus $Z=(-c_1/c_2)^{1/(-\alpha+1)}$
with $c_1$, $c_2$ and $\alpha$ fitted to the cumulative distribution data~\cite{girardiAva}.
At the critical point, the cutoff is expected to scale as $Z\sim L^D$ and $D$ is an exponent
defining a characteristic dimensionality of the avalanches~\cite{pruessnerSOC2012}.
If this scaling relation does not hold, then the system is not critical.

%


\begin{figure*}[t!]
	\includegraphics[width=170mm]{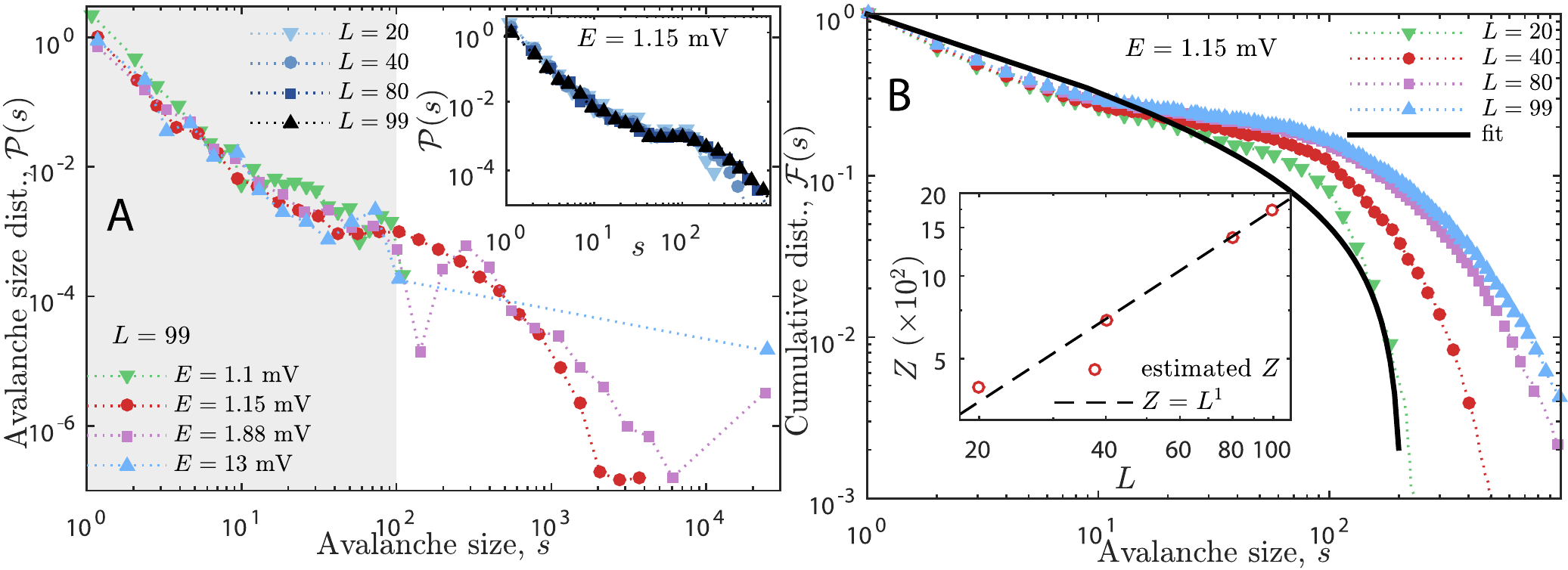}
	\caption[Avalanche size distributions and cumulative distributions.]{\label{fig:avalDist}
	Avalanche size distributions and cumulative distributions.
	Panel~A: Typical avalanche distributions $\mathcal{P}(s)$ for many $E$:
    (\textcolor{greenC}{$\blacktriangledown$}) inactive phase,
    (\textcolor{redC}{\ding{108}}) critical phase,
    (\textcolor{purpleC}{$\blacksquare$}) and (\textcolor{blueC}{$\blacktriangle$}) percolating phases.
    Note the bump around avalanches of size $N_c\approx200$ for $E=1.88\mV$ (\textcolor{purpleC}{$\blacksquare$}); this bump reveals
    the internal structure of the network (see text for discussion). The light gray background highlights the range of
    avalanche size $s$ in which all phases have PL-shaped distributions.
    Panel~A inset: avalanche distributions for the critical phase ($E=1.15\mV$) for increasing $L$.
    Panel~B: avalanche cumulative distributions $\mathcal{F}(s)$ corresponding to panel~A inset, $E=1.15\mV$ (the critical phase),
    for increasing $L$. Solid line is the fit of Equation~\eqref{eq:cDistFit} used to estimate $Z(L=20)$ and $\alpha=1.4(1)$.
    Panel~B inset: scaling law of $Z\sim L^D$ for $E=1.15\mV$ yielding $D=1.0(1)$; this scaling holds inside the whole
    critical phase. Avalanche distributions for the other phases have different $D$ and are presented in Figure~\ref{fig:collapse}.}
\end{figure*}
\begin{figure*}[t!]
	\includegraphics[width=170mm]{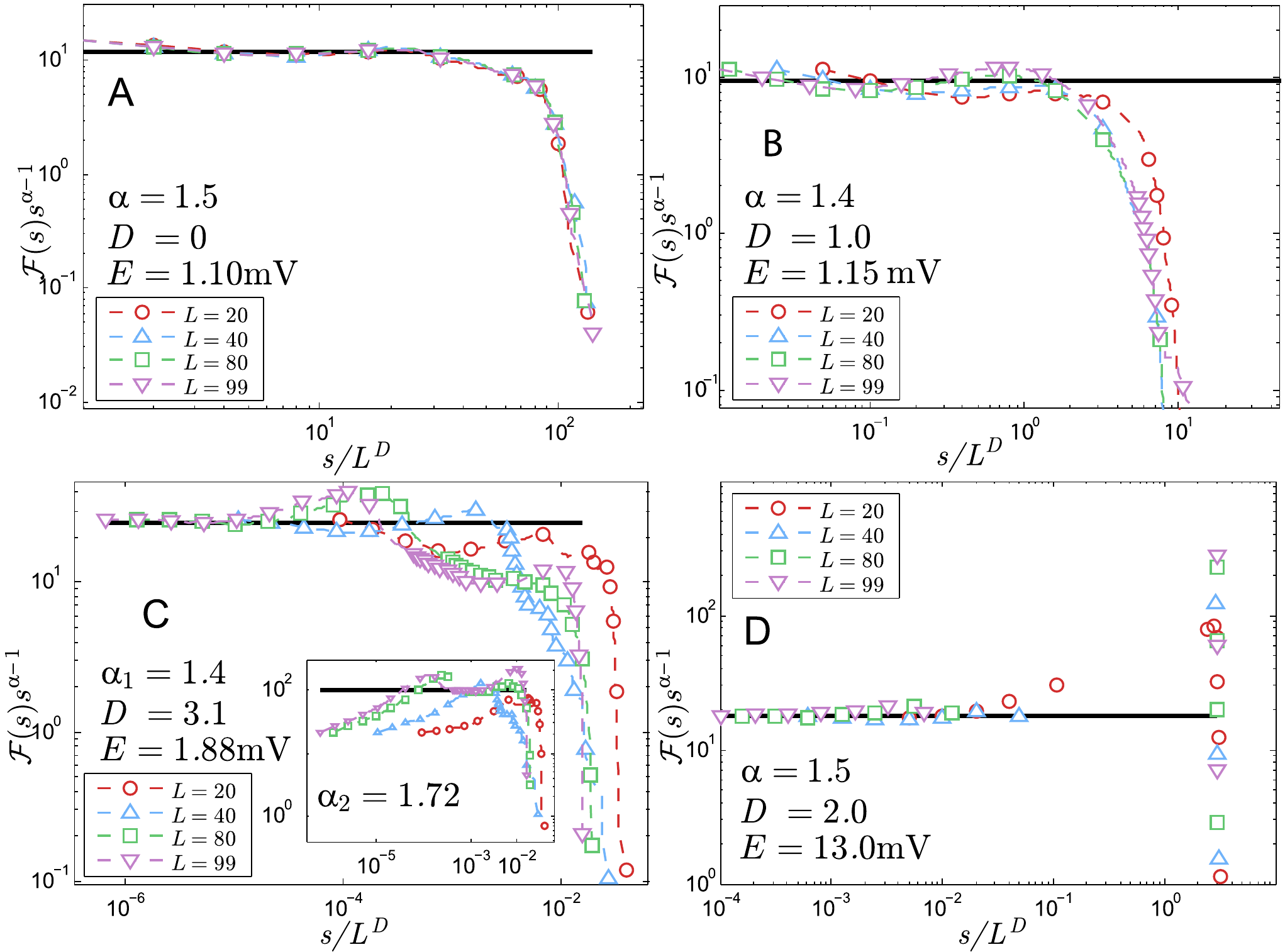}
	\caption[Collapse of avalanche size cumulative distributions.]{\label{fig:collapse}
	Collapse of avalanche size cumulative distributions for many realizations of the simulation for each $E$.
	Panel~A: Collapse of avalanches in inactive phase; the cutoff does not scale with system size ($D=0$) but
    the distribution presents a	PL regime with exponent $\alpha=1.5(1)$ due to noisy activity in the \camLGN.
	Panel~B: Collapse of avalanches in Griffiths (critical) phase with exponents $\alpha=1.4(1)$ and $D=1.0(1)$
    corresponding to Figure~\ref{fig:avalDist}B.
	Panel~C: Collapse of avalanches in weakly percolating phase
	with exponents $\alpha_1=1.4(1)$, $\alpha_2=1.72(8)$ and $D=3.1(3)$; the bump separating both PL ranges is a consequence of the
    columnar structure of the network, as it lies where $s\approx N_c$.
	Panel~D: Collapse of avalanches in strongly percolating phase with exponents $\alpha=1.5(1)$ and $D=2$;
    the gap in this distribution shows that propagation occurs mainly
	through a large dominating avalanche, so the PL scaling represents only noisy avalanche activity in the \camLGN.
	The straight black lines guide the eyes over the collapse of the PLs.
    Notice that $\alpha$ is the exponent of $\mathcal{P}(s)$.
    All the PL exponents were checked using Maximum Likelihood test (see Supplementary Material).}
\end{figure*}
\begin{figure*}[t!]
	\includegraphics[width=170mm]{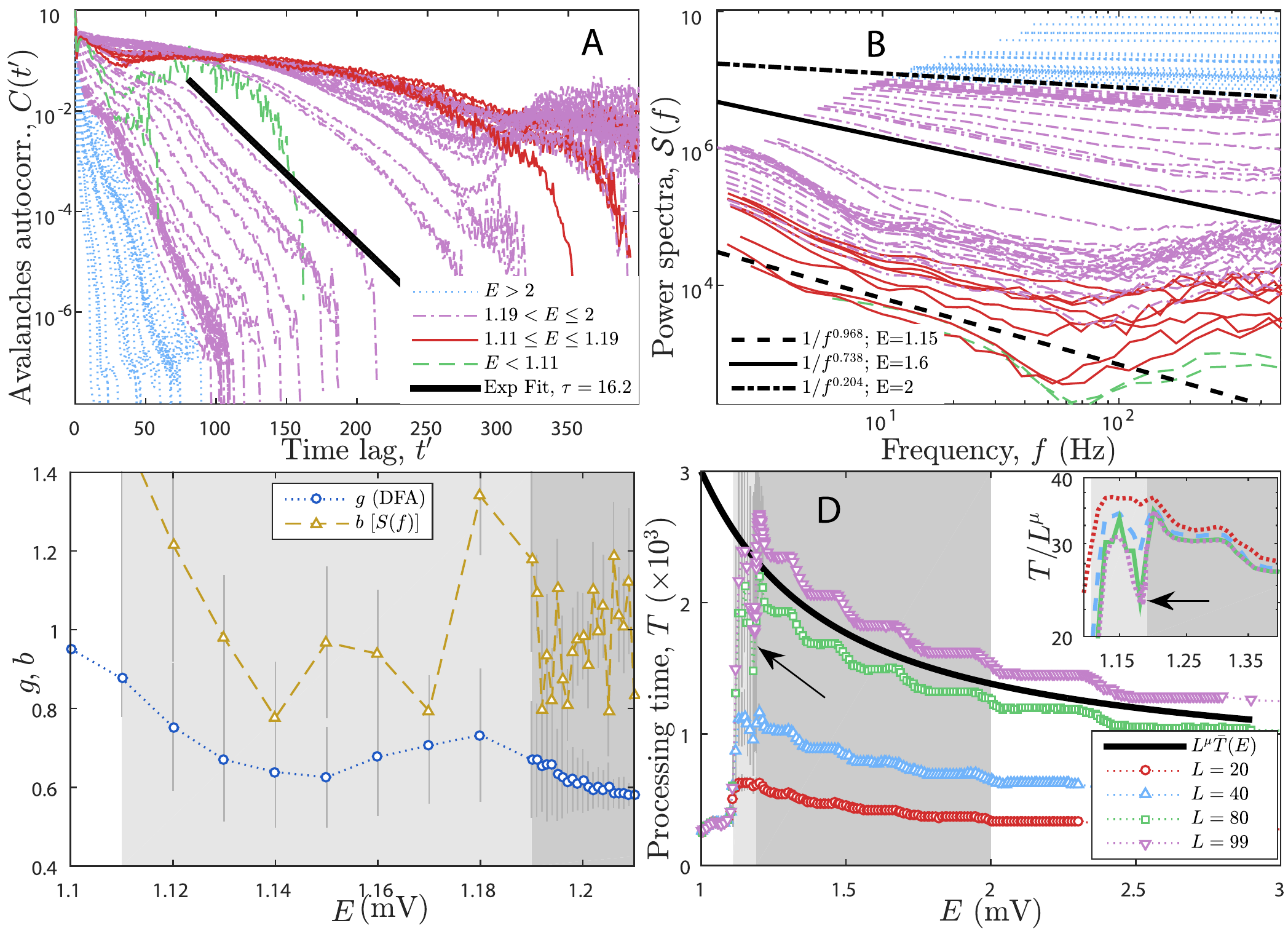}
	\caption[Autocorrelation, power spectrum, DFA and processing time.]{\label{fig:corrpowexp}Autocorrelation, power spectrum,
    DFA and processing time.
	Panel~A,B: Average autocorrelation (A) and average power spectrum (B) of avalanche sizes time series
	for $L=99$ and many $E$: 
    (\textcolor{greenC}{$---$}) inactive phase,
    (\textcolor{redC}{------}) critical phase,
    (\textcolor{purpleC}{$-\cdot-$}) and (\textcolor{blueC}{$\cdots\cdot\cdot$}) percolating phases.
    The bold lines show the exponential cutoff fit of the autocorrelation $E=1.6\mV$ giving $\tau=16.2$~ts (A).
	and the fit $S(f)\sim f^{-b}$ giving $b=0.97(5)$ for $E=1.15\mV$ (critical phase),
    $b=0.74(4)$ for $E=1.6\mV$ and $b=0.20(2)$ for $E=2\mV$ (weakly percolating phase) as examples (B).
	Notice how the curves' slope in panel B smoothly vary and yield $b\sim1$ inside the Griffiths phase (see panel C too).
	Panel~C: Power spectrum and DFA exponents versus $E$ inside and close to the Griffiths region.
	$b$ is the avalanches power spectrum exponent and $g$ is the DFA exponent of the activity time series (of Figure~\ref{fig:tsOrderSusc}A).
    Notice that $b\approx1$ and $g\gtrsim0.7$ inside the critical region indicating long-range temporal correlations of both the avalanche
    time series and the spiking activity of the network.
    Panel~D: Processing time $T$ as function of $E$ for many $L$; solid line indicates the asymptotic behavior of $T$.
	Panel~D inset: collapse plot $\bar{T}=T/L^{\mu}$ yielding $\mu=0.9(1)$.
	Arrows indicate the minimum of $T$, vertical bars are standard deviation,
    light gray is the Griffiths phase and dark gray is the weakly percolating phase. See Supplementary Material for
    more details about $T$.}
\end{figure*}

\subsection*{Mean-field approximation}

It is clear in Fig.~\ref{fig:tsOrderSusc}A that there is a change
in the activity profile as $E$ is varied.
The signal may reach only a few neurons of each layer for $E=1.1\mV$
whereas for $E\gtrsim1.19\mV$, it excites the whole network.
Such a behavior arises from the competition between
excitation and dissipation. The excitation level of the network is controlled by the EPSP parameter, $E$,
whereas the dissipation level is controlled by the attenuation constant, $\lambda$.

At the point where excitation balances dissipation, one
expects that every neuron in the network will fire once.
We will call this point the \textit{activation threshold},
where $E=E_{th}$.
In order for any neuron $i$ to fire, the signal that reaches
the soma should be greater than or equal to the firing threshold, $v_T$:
\begin{equation}
\label{eq:mfApprox}
n_{in}^{(i)}\langle d_{100}^{(i)}\rangle\geq v_T\ \virgula
\end{equation}
where $n_{in}^{(i)}$ is the average amount of inward synapses 
per dendritic compartment of the postsynaptic neuron~$i$,
$\langle d_{100}^{(i)}\rangle=r_{100}^{(i)}(\lambda)E$ is the
average signal arriving on the soma compartment
from one single synapse on any dendritic compartment of neuron~$i$ and
$r_{100}^{(i)}(\lambda)=\lambda^{50(1+\ln\lambda)}$
is the fraction of the signal that reaches
the soma as function of $\lambda$ from a single presynaptic cell.
See Supplementary Material for a
derivation of~$\langle d_{100}^{(i)}\rangle$.

Since the structure of the network is fixed for each trial, we may estimate
$n_{in}^{(i)}\approx6.5$ (from Table~S2 in Supplementary Material).
It is then simple to calculate $E_{th}$ from Equation~\eqref{eq:mfApprox}:
$E_{th}=v_T/[n_{in}^{(i)}r_{100}^{(i)}(\lambda)]\approx1.88\mV$,
which we know from numerical data to be overestimated.
Anyway, it is still close to the average experimental value of the $E$ in the
cortex~\cite{williamsV12002,songEPSP2005,lefortEPSP2009}.


\subsection*{Computational results}

We performed between one hundred and three hundred trials of the network disorder
for each EPSP of each $L$ in order to calculate averages [equations~\eqref{eq:activedens} to \eqref{eq:proctime}] and
standard deviations (vertical bars in Figure~\ref{fig:tsOrderSusc}).

\subsubsection*{Griffiths phase and critical phase transition}


A critical point is found for $E=E_c$
if~\cite{stanley,barberFSS1983,noneqPhaseTrans2008,vojtaGrifRev2006,munozGrif2013}:
(a) $\rho(E=E_c;L)\rightarrow0$ according to Equation~\eqref{eq:activedens}, and
(b) $\chi_{\rho}(E=E_c;L)\rightarrow+\infty$ according to Equation~\eqref{eq:suscept}; both when $L\rightarrow\infty$.
It is easy to notice that both conditions are fulfilled within the light gray region in Figure~\ref{fig:tsOrderSusc}B
(circles stand for $\rho$ and squares stand for $\chi_{\rho}$).
The light gray region is called a Griffiths phase because it is an extended region that satisfies both conditions,
instead of being a single critical point.
This Griffiths phase is due to the rewiring of the network, which originates rare regions. In some connection patterns the network
allows neurons to fire more easily. Although rare, the presence of this behavior in the range of $1.11\leq E\leq1.19$ causes large
fluctuation of $\rho$ when several trials of the network are considered.

In fact, Figures~\ref{fig:tsOrderSusc}C,D show that the PL scale functions [Equations~\eqref{eq:activedens} and~\eqref{eq:suscept}]
are satisfied inside the whole EPSP range $1.11\leq E\leq1.19$: the circles and squares in the light gray region in
Figure~\ref{fig:tsOrderSusc}B correspond to the red circles in panels~C and~D.
These panels explicitly show that $\rho\rightarrow0$ and $\chi_{\rho}\rightarrow\infty$ for $L\rightarrow\infty$.
We fitted these equations to the data on $E=1.19\mV\equiv E_c$
and obtained the critical exponents~$\beta/\nu_{\bot}=0.55(3)$, $\gamma/\nu_{\bot}=3.1(2)$ and
$\gamma'/\nu_{\bot}=(\gamma-\beta)/\nu_{\bot}=2.5(2)$.
See the Supplementary Material for comments on the values of these exponents.
Notice that $E_c=1.19\mV$ is of the same order of our naive mean-field
calculation, $E_{th}\approx1.88$, although $E_c$ is even closer to the average EPSP in
the cortex~\cite{williamsV12002,songEPSP2005,lefortEPSP2009}.

The left-hand side of the Griffiths phase (white background in Figure~\ref{fig:tsOrderSusc}B)
has $\rho=0$ (green upside down triangles
in Figures~\ref{fig:tsOrderSusc}C,D) and is named the inactive phase whereas the right-hand side (dark grey background)
of the Griffiths has $\rho$ growing rapidly to saturated $\rho\approx1$ (purple squares and blue triangles
in Figures~\ref{fig:tsOrderSusc}C,D) and is named the percolating phase. Both of these phases have finite susceptibility
for increasing $L$.


\subsubsection*{Avalanche distributions}

PL avalanche size and duration distributions are believed to be the ultimate signature of criticality
in avalanching dynamical systems since the seminal work of Bak~et~al.~\cite{bakPRL,pruessnerSOC2012}.
However, recent works have shown that either critical systems may have no PL distributed
avalanches~\cite{hartleyISuscep2013} or non-critical systems may have
PL distributed avalanches~\cite{destexhePL2015}. We have shown in the previous section that our model has a continuous phase
transition through a GP. Then, this section is devoted to show that PL avalanches may emerge in our system even outside of the
critical regime. The scaling of the cutoff of the distributions must be calculated as function of the system size in order
to have a better estimate of the critical regime of the system. In addition, we highlight some features of the
avalanche distributions which are related to the structure of the network.

Figure~\ref{fig:avalDist}A shows the distribution of avalanche sizes $s$ for $L=99$
and four typical EPSP: $E=1.1\mV$ (inactive phase), $E=1.15\mV$ (Griffiths phase), $E=1.88\mV$ (percolating phase),
and $E=13\mV$ (percolating phase). Notice that all the four distributions have PL shape inside the highlighted region
($s\leq100$). We confirmed the PL decay of these distributions via the Maximum Likelihood test presented in the Supplementary Material.
Therefore, a PL shape alone in the distribution of avalanche sizes is not enough to determine which distribution correspond to a critical
regime. Sethna's scaling law (a law which relates avalanche sizes and durations and is expected to hold only in criticality) also holds
for the four phases of the model~\cite{girardiV1conf2015}.

The distribution of avalanche sizes for fixed $E=1.15\mV$ (inside the GP) is shown in the inset of Figure~\ref{fig:avalDist}A for
systems of sizes $L=20,40,80,99$. Notice that the larger the system the farther the reach of the distribution.
This fact becomes clearer in the plot of the cumulative distribution of $s$ in Figure~\ref{fig:avalDist}B.
Although the cumulative distributions seem to be turning flat for large $L$, they are actually accumulating towards a non-zero slope
as $L$ increases corresponding to a PL with exponent $-\alpha+1$.
We fitted Equation~\eqref{eq:cDistFit} to the distributions in this figure in order to estimate their cutoffs, $Z(L)$,
and their PL exponent, $\alpha=1.4(1)$ (this value is obtained for all system sizes).
The plot of $Z(L)$ is in the inset of Figure~\ref{fig:avalDist}B. We fitted
$Z(L)\sim L^D$ for the cutoffs and obtained $D=1.0(1)$. This value agrees with the scaling exponent for the largest avalanche of
the system presented in the Supplementary Material.

Figure~\ref{fig:collapse} shows the collapse of the cumulative distributions, $s^{\alpha-1}\mathcal{F}(s/L^D)$ for many $L$.
Each panel corresponds to a different phase of the model. Panel B of Figure~\ref{fig:collapse} corresponds to the collapsed
data of Figure~\ref{fig:avalDist}B yielding $\alpha=1.4$ and $D=1.0(1)$, evidencing the PL shape and the scaling of the cutoff
of the cumulative distribution in the GP. Both values agree very well with the fitted data from Figure~\ref{fig:avalDist}B.
The data corresponding to the inactive (or subcritical) phase is collapsed in Figure~\ref{fig:collapse}A yielding $D=0$
(avalanches do not scale with system size) and $\alpha=1.5$.
The data corresponding to the percolating (or supercritical) phase is presented in panels C and D of Figure~\ref{fig:collapse}.
$D$ varies inside this phase between $D=3$ (for $1.2\lesssim E\lesssim2\mV$) and $D=2$ and becomes $D=2$ for large $E$.
$\alpha$ also varies and becomes $\alpha=1.5$ for large $E$.

The characteristic dimension $D$ of avalanches sizes indicates how the activity spreads throughout the network~\cite{girardiV1conf2015}:
$D=1$ means that the activity is rather spreading through the columns of the model, $D=3$ means that activity is spreading both radially and
inside the columns and $D=2$ means that the activity is spreading only radially and simultaneously within all the layers.
Thus, the different values of $D$ in the percolating phase gives rise to two dynamically distinct phases:
the weakly percolating regime (for $1.2\lesssim E\lesssim2\mV$) and the strongly percolating regime (for $E\gtrsim2\mV$).
A detailed discussion concerning the spreading of the network activity is presented elsewhere~\cite{girardiV1conf2015}.

We identified two PL ranges for the avalanche distributions in the weakly percolating regime (see the purple squares
in Figure~\ref{fig:avalDist}A and all the curves in Figure~\ref{fig:collapse}C). The first range, $s<N_c$,
has $\alpha_1=1.4(1)$ and the second range, $s>N_c$, has $\alpha_2=1.7(1)$. $N_c\approx200$ is the
amount of neurons in a single column of the network. These ranges are separated by a characteristic bump
located at $s\approx N_c$ that is generated by the columnar structure of the network. If the column size, $l$,
tended to $L$, the left-most bump would move to the right until merging with the right-most bump
(the cutoff of the distribution).
Then, there would be a single PL for the avalanches as the one presented by
the columnless layered model of Teramae~\& Fukai~\cite{teramaeFukai2007}.


\subsubsection*{Long-range correlations and processing time}

From Figure~\ref{fig:tsOrderSusc}A, it is clear that avalanches are temporally organized: each of the
four presented time series has a low amplitude beginning, a growth until a maximum amplitude and then the activity
the activity decreases until fading out. We computed the sequence of avalanche sizes, $s(n)$, its autocorrelation, $C(t')$,
and its power spectrum, $S(f)$. The time $t'$ is the lag between every two avalanches, $s(n)$ and $s(n-t')$.
The DFA is calculated over the raw time series, $A(t)$, presented in Figure~\ref{fig:tsOrderSusc}A.
These calculations are described in Supplementary Material.

The autocorrelation function of avalanche sizes is presented in Figure~\ref{fig:corrpowexp}A and
the power spectrum of avalanche sizes is given in Figure~\ref{fig:corrpowexp}B.
Each of the curves is averaged out of many realizations of the simulation.
The more long-lasting correlations are inside the Griffiths phase (red curves) or very close to it
(purple curves for $E\gtrsim1.19\mV$). The characteristic time $\tau$ has been
fitted by the exponential cutoff of Equation~\eqref{eq:corrFunc}.
$\tau$ is expected to scale with $\exp(L^D)$ inside a Griffiths phase~\cite{brayGrif1988,vojtaGrifRev2006,munozGrif2013}.
We present the study of $\tau$ in the Supplementary Material.

The power spectrum shows a stable $f^{-b}$ behavior with $b\approx1$
inside the critical phase (see Figure~\ref{fig:corrpowexp}B, red curves) for $f<100~\textnormal{Hz}$,
suggesting long-range correlations in the avalanche size time series.
In fact, the smooth change of slopes in Figure~\ref{fig:corrpowexp}B indicates that $b$ varies continuously with $E$
(green squares in Figure~\ref{fig:corrpowexp}C).
We calculated the DFA exponent $g$ for the activity time series in order to verify the long-range correlation.
The exponent $g$ also varies continuously with $E$ (purple upside down triangles in Figure~\ref{fig:corrpowexp}C)
and remains bounded in the interval $0.7\leq g\leq1$
inside the Griffiths phase, confirming the presence of long-range correlations.
Both exponents $b$ and $g$ are close to the expected experimental values~\cite{novikov1fB1997,henrie1fV12005,hermes1fV12014}.

Figure~\ref{fig:tsOrderSusc}A also shows that the activity propagation time (also known as mean survival time in absorbing state
phase transitions~\cite{noneqPhaseTrans2008}) has a non-monotonic behavior with $E$.
Figure~\ref{fig:corrpowexp}D shows in details how $T$ varies with $E$.
We found that $T$ has a deep local minimum at $E=1.18\mV$ (pointed by arrows) and a global maximum
around $E=1.2\mV$. The maximum grows with system size in the critical point $E_c$ according to the law $T\sim L^{\mu}$
[Equation \eqref{eq:proctime}], as expected. After that, $T$ slowly decays
asymptotically through a landscape full of shallow minima.
The more intense the EPSP, the less active presynaptic neurons are needed to propagate the signal.
As a result, the network as a whole will take less time to get activated for large $E$.
The solid line is an arbitrary fit to the asymptotic decay of $T(E)$.
The FSS exponent of the propagation time is $\mu=0.90(1)$ [see Equation~\eqref{eq:proctime} and Supplementary Material].
The rescaled processing time, $\bar{T}\sim T/L^{\mu}$, is shown in the inset of Figure~\ref{fig:corrpowexp}D.

For $E<E_c$, $T$ versus $E$ is shaped similarly to $\rho$ (compare dark circles in Figure \ref{fig:tsOrderSusc}B
to the behavior of $T$ in Figure \ref{fig:corrpowexp}D): the more neurons get activated, the more time the signal takes to
completely cease. Such proportionality causes a deep local minimum in $T$ near the phase transition and inside the critical phase.
Additionally, $T$ has maximum variance inside all the Griffiths region (see Supplementary Material),
meaning that although the signal propagates quickly (in average), the system is flexible enough
to adapt to external input as suggested by experimental work~\cite{shewV1Aval2015}.
Therefore, it may be conjectured that the processing of information occur preferably around this region.


\section*{Concluding remarks}

We studied a model for the visual cortex presenting its
characteristic columnar structure. The key elements of this model are the
neurons dynamical structure and dendritic excitation/dissipation balance. While the delay
due to action potential propagation in the dendrites and axons causes the
avalanches separation of time scales, the signal attenuation balances the network information
processing. If we were to model lateral inhibition inside the layers,
we would expect the activation threshold $E_{th}$ and the critical point $E_c$ to grow in accordance with inhibition levels,
leaving the described phase transition qualitatively unaltered.
Certainly, the absorbing states of the system would depend on the reach, the quantity and the intensity of the
lateral connections inside the layers.
The detailed analysis of this scenario is a matter of future work.

We may summarize this work within five main findings:
\begin{itemize}
    \item The extended critical phase is confirmed by the scaling laws of the density of activated neurons
    and of its associated susceptibility.
    The quenched disorder of the network generates the zero mean and the divergent
    variance of the density of activated neurons in an extended region of the
    parameter space.
    This phase is known as a Griffiths phase~\cite{vojtaGrifRev2006,munozGrif2013}.
    The critical phase lies near the average value of EPSP in the cortex
    (namely $E\approx1\mV$~\cite{williamsV12002,songEPSP2005,lefortEPSP2009}).
    \item Avalanche size distributions are PL-shaped and their cutoff scale with system
    size as expected inside the critical region. However, avalanches also present a PL shape within
    a limited range of $s$ for non-critical phases, corroborating the argument that PL
    avalanches alone (without the cutoff scaling law verified) are not necessarily
    connected to criticality~\cite{hartleyISuscep2013,destexhePL2015}.
    \item The model displays long-range temporal correlations inside the critical region:
    the power spectrum of avalanche sizes have the form $1/f^{b}$ with $b\approx1$ whereas the activity time series
    DFA exponent is $g\gtrsim0.7$;
    both match experimental results~\cite{novikov1fB1997,linkenkaerSOC,henrie1fV12005,hermes1fV12014}.
    \item The model columnar structure was
evidenced in the characteristic scale of small avalanches in the avalanches
size distributions for the weakly ordered regime.
We hypothesize that the two different power-law regimes could be experimentally found if the visual system was
working in a slightly supercritical regime; for instance, with $E=1.88\mV$.
    \item We discovered a local minimum of the network processing time inside the 
    Griffiths phase close to $E_c$. At $E=E_c$, the propagation time diverges
    as expected for absorbing phase transitions.
    Although minimum, the variance of the propagation time is maximum, resembling the
    variance of the order parameter.
    We conjecture that this behavior could be essential for a reliable processing of
    information. Inside the inactive or strongly percolating phases, either the network
    will respond with noisy activity or will fire a quick single dominating avalanche.
\end{itemize}
All these evidences strongly suggests that visual processing occurs preferably at or near
the critical phase. Our system's observed criticality results from the average over the
rewiring of the network quenched disorder. In the real brain, cortex synapses are reinforced
or weakened on the time scales of seconds to hours, altering the microscopic structure
of the network~\cite{chungRew2002}. This  phenomenon is known as sensory
adaptation~\cite{alamancosRew2004}. Moreover, the rewiring of brain connections
occurs preferably during quiescent states, in which there is no processing
of information~\cite{alamancosRew2004} -- similarly, after the network has become inactive, 
we rewire it in order to stimulate the system again .
Thus, Griffiths phases could potentially be experimentally measured if one considers a
long-term measurement of brain activity. Such data would naturally comprise different
trials of connections for the propagation of activity in the many regions of the cortex.
Therefore, averages of the long-term activity could potentially be subject to rare region
effects.

The studied model presents essential SOC features, such as
the separation of time scales (which emerges naturally in the critical regime),
PL avalanches that scale with system size, long-range temporal
correlations and approximately $1/f$ power spectrum (inside the critical regime).
Still, this model does not present an explicit mechanism of self-organization
that would allow it to reach the critical state independently of its parameters.
Anyhow, the real visual system may adapt to external stimuli and is indeed
self-organized~\cite{chungRew2002,shewV1Aval2015}. Besides,
our model presents an extended region of critical behavior and gives us hints that
these rare region effects might play a role in the real systems. If so, the critical
state would be easier to achieve by evolutionary means due to its extended
region in the parameter space.

It is not our aim in this work to determine the universality class of our model. Some issues need to be addressed in order to do that:
\begin{enumerate}
    \item The dimensionality of our network is not trivially determined: although $N\sim L^2$, the connections are chosen randomly between
    neurons of adjacent layers within a very small and well determined excitatory field. Notice also that multiple interactions between
    the same neighbors happen. On the other hand, mean-field (random) networks have connections distributed all around the network without
    multiple interactions.
    \item Our exponents $\alpha$ and $D$ are well determined but they rely on an unusual definition of avalanches: we define avalanches
    between two consecutive instants of soma silence whereas usually they are defined between two absorbing states of the system.
    \item The dynamics of our model is somewhat similar to the general epidemic process (GEP) or generalized GEP~\cite{janssenEPDynP2004};
    however, $\gamma'$ is defined for our model but not for these systems.
\end{enumerate}

We stress that the strong way to describe critical phase transitions in the brain
is to define an order parameter and its associated susceptibility and check in which range
of parameters the first goes continuously to zero and the second diverges according to power laws, instead of
only studying avalanche distributions or long-range correlations.

Some future work may focus on improving some of the model's features. We can add disorder in $E$, background noise, synaptic dynamics or plasticity~\cite{levina,ariadneDynSyn2015,shewV1Aval2015} to model adaptability,  and use
different input stimuli. We also plan to make the excitatory field of each neuron to change with depth

Our study indicates that being critical or quasi-critical is advantageous for the brain sensory networks.
Our network architecture could be further used to inspire the development of pattern recognition applications
because of the short processing time inside the critical region. The high variability of activated density of neurons
may enhance the sensitivity to different patterns, which may also aid in the pattern recognition tasks.
Finally, we hope to provide here a kinematic framework for microscopic cortical modeling.


\begin{thebibliography}{10}
\expandafter\ifx\csname url\endcsname\relax
  \def\url#1{\texttt{#1}}\fi
\expandafter\ifx\csname urlprefix\endcsname\relax\def\urlprefix{DOI }\fi
\providecommand{\bibinfo}[2]{#2}
\providecommand{\eprint}[2][]{\url{#2}}

\bibitem{chialvoArx2008}
\bibinfo{author}{Chialvo, D.~R.}, \bibinfo{author}{Balenzuela, P.} \&
  \bibinfo{author}{Fraiman, D.}
\newblock \bibinfo{title}{The brain: What is critical about it?}
\newblock \emph{\bibinfo{journal}{AIP Conf. Proc.}}
  \textbf{\bibinfo{volume}{1028}}, \bibinfo{pages}{28} \urlprefix\url{10.1063/1.2965095} (\bibinfo{year}{2008}).

\bibitem{hidalgoEvolSOC2014}
\bibinfo{author}{Hidalgo, J.} \emph{et~al.}
\newblock \bibinfo{title}{Information-based fitness and the emergence of
  criticality in living systems}.
\newblock \emph{\bibinfo{journal}{Proc. Nat. Acad. Sci. (USA)}}
  \textbf{\bibinfo{volume}{111(28)}}, \bibinfo{pages}{10095--10100}
  \urlprefix\url{10.1073/pnas.1319166111} (\bibinfo{year}{2014}).

\bibitem{turing1950}
\bibinfo{author}{Turing, A.~M.}
\newblock \bibinfo{title}{Computing machines and intelligence}.
\newblock \emph{\bibinfo{journal}{Mind}} \textbf{\bibinfo{volume}{59}},
  \bibinfo{pages}{433--460} (\bibinfo{year}{1950}).

\bibitem{chialvoReview}
\bibinfo{author}{Chialvo, D.~R.}
\newblock \bibinfo{title}{Emergent complex neural dynamics}.
\newblock \emph{\bibinfo{journal}{Nat. Phys.}} \textbf{\bibinfo{volume}{6}},
  \bibinfo{pages}{744--750} (\bibinfo{year}{2010}).

\bibitem{plenzBenefits}
\bibinfo{author}{Shew, W.~L.} \& \bibinfo{author}{Plenz, D.}
\newblock \bibinfo{title}{The functional benefits of criticality in the
  cortex}.
\newblock \emph{\bibinfo{journal}{Neuroscientist}}
  \textbf{\bibinfo{volume}{19(1)}}, \bibinfo{pages}{88--100}
  \urlprefix\url{10.1177/1073858412445487} (\bibinfo{year}{2013}).

\bibitem{beggsEditorial}
\bibinfo{author}{Beggs, J.}
\newblock \bibinfo{title}{Editorial: Can there be a physics of the brain?}
\newblock \emph{\bibinfo{journal}{Phys. Rev. Lett.}}
  \textbf{\bibinfo{volume}{114(22)}}, \bibinfo{pages}{220001}
  \urlprefix\url{10.1103/PhysRevLett.114.220001} (\bibinfo{year}{2015}).

\bibitem{bakPRL}
\bibinfo{author}{Bak, P.}, \bibinfo{author}{Tang, C.} \&
  \bibinfo{author}{Wiesenfeld, K.}
\newblock \bibinfo{title}{Self-organized criticality: An explanation of 1/f
  noise}.
\newblock \emph{\bibinfo{journal}{Phys. Rev. Lett.}}
  \textbf{\bibinfo{volume}{59(4)}}, \bibinfo{pages}{381--384}
  (\bibinfo{year}{1987}).

\bibitem{kinouchiCopelli}
\bibinfo{author}{Kinouchi, O.} \& \bibinfo{author}{Copelli, M.}
\newblock \bibinfo{title}{Optimal dynamical range of excitable networks at
  criticality}.
\newblock \emph{\bibinfo{journal}{Nat. Phys.}} \textbf{\bibinfo{volume}{2}},
  \bibinfo{pages}{348--351} (\bibinfo{year}{2006}).

\bibitem{socPlasticity}
\bibinfo{author}{de~Arcangelis, L.}, \bibinfo{author}{Perrone-Capano, C.} \&
  \bibinfo{author}{Herrmann, H.~J.}
\newblock \bibinfo{title}{Self-organized criticality model for brain
  plasticity}.
\newblock \emph{\bibinfo{journal}{Phys. Rev. Lett.}}
  \textbf{\bibinfo{volume}{96}}, \bibinfo{pages}{028107}
  (\bibinfo{year}{2006}).

\bibitem{beggsCritical}
\bibinfo{author}{Beggs, J.~M.} \& \bibinfo{author}{Timme, N.}
\newblock \bibinfo{title}{Being critical of criticality in the brain}.
\newblock \emph{\bibinfo{journal}{Front. Physiol.}}
  \textbf{\bibinfo{volume}{3}}, \bibinfo{pages}{163} (\bibinfo{year}{2012}).

\bibitem{mosqueiro2013}
\bibinfo{author}{Mosqueiro, T.~S.} \& \bibinfo{author}{Maia, L.~P.}
\newblock \bibinfo{title}{Optimal channel efficiency in a sensory network}.
\newblock \emph{\bibinfo{journal}{Phys. Rev. E}}
  \textbf{\bibinfo{volume}{88(1)}}, \bibinfo{pages}{012712}
  \urlprefix\url{10.1103/PhysRevE.88.012712} (\bibinfo{year}{2013}).

\bibitem{beggsPlenz2003}
\bibinfo{author}{Beggs, J.~M.} \& \bibinfo{author}{Plenz, D.}
\newblock \bibinfo{title}{Neuronal avalanches in neocortical circuits}.
\newblock \emph{\bibinfo{journal}{J. Neurosci.}}
  \textbf{\bibinfo{volume}{23(35)}}, \bibinfo{pages}{11167--11177}
  (\bibinfo{year}{2003}).

\bibitem{ribeiroCopelli}
\bibinfo{author}{Ribeiro, T.~L.} \emph{et~al.}
\newblock \bibinfo{title}{Spike avalanches exhibit universal dynamics across
  the sleep-wake cycle}.
\newblock \emph{\bibinfo{journal}{PLoS ONE}} \textbf{\bibinfo{volume}{5(11)}},
  \bibinfo{pages}{e14129} (\bibinfo{year}{2010}).

\bibitem{plenzAvalV12010}
\bibinfo{author}{Hahn, G.} \emph{et~al.}
\newblock \bibinfo{title}{Neuronal avalanches in spontaneous activity in vivo}.
\newblock \emph{\bibinfo{journal}{Journal of Neurophysiology}}
  \textbf{\bibinfo{volume}{104(6)}}, \bibinfo{pages}{3312--3322}
  \urlprefix\url{10.1152/jn.00953.2009} (\bibinfo{year}{2010}).

\bibitem{priesemannSub2014}
\bibinfo{author}{Priesemann, V.} \emph{et~al.}
\newblock \bibinfo{title}{Spike avalanches in vivo suggest a driven, slightly
  subcritical brain state}.
\newblock \emph{\bibinfo{journal}{Front Syst Neurosci.}}
  \textbf{\bibinfo{volume}{8}}, \bibinfo{pages}{108} \urlprefix\url{10.3389/fnsys.2014.00108} (\bibinfo{year}{2014}).

\bibitem{shewV1Aval2015}
\bibinfo{author}{Shew, W.~L.} \emph{et~al.}
\newblock \bibinfo{title}{Adaptation to sensory input tunes visual cortex to
  criticality}.
\newblock \emph{\bibinfo{journal}{Nat. Phys.}} \textbf{\bibinfo{volume}{11}},
  \bibinfo{pages}{659--663} \urlprefix\url{10.1038/nphys3370} (\bibinfo{year}{2015}).

\bibitem{linkenkaerSOC}
\bibinfo{author}{Linkenkaer{-}Hansen, K.}, \bibinfo{author}{Nikouline, V.~V.},
  \bibinfo{author}{Palva, J.~M.} \& \bibinfo{author}{Ilmoniemi, R.~J.}
\newblock \bibinfo{title}{Long-range temporal correlations and scaling behavior
  in human brain oscillations}.
\newblock \emph{\bibinfo{journal}{J. Neurosci.}}
  \textbf{\bibinfo{volume}{21(4)}}, \bibinfo{pages}{1370--1377}
  (\bibinfo{year}{2001}).

\bibitem{haimoviciRestState2013}
\bibinfo{author}{Haimovici, A.}, \bibinfo{author}{Tagliazucchi, E.},
  \bibinfo{author}{Balenzuela, P.} \& \bibinfo{author}{Chialvo, D.~R.}
\newblock \bibinfo{title}{Brain organization into resting state networks
  emerges at criticality on a model of the human connectome}.
\newblock \emph{\bibinfo{journal}{Phys. Rev. Lett.}}
  \textbf{\bibinfo{volume}{110}}, \bibinfo{pages}{178101}
  (\bibinfo{year}{2013}).

\bibitem{barberFSS1983}
\bibinfo{author}{Barber, M.~N.}
\newblock \bibinfo{title}{Finite-size scaling}.
\newblock In \bibinfo{editor}{Domb, C.} \& \bibinfo{editor}{Lebowitz, J.~L.}
  (eds.) \emph{\bibinfo{booktitle}{Phase Transitions and Critical Phenomena}},
  vol.~\bibinfo{volume}{8} (\bibinfo{publisher}{Academic Press},
  \bibinfo{address}{New York, USA}, \bibinfo{year}{1983}).

\bibitem{stanley}
\bibinfo{author}{Stanley, H.~E.}
\newblock \emph{\bibinfo{title}{Introduction to Phase Transitions and Critical
  Phenomena}} (\bibinfo{publisher}{Oxford University Press},
  \bibinfo{address}{New York, UK}, \bibinfo{year}{1971}).

\bibitem{noneqPhaseTrans2008}
\bibinfo{author}{Henkel, M.}, \bibinfo{author}{Hinrichsen, H.} \&
  \bibinfo{author}{L{\"u}beck, S.}
\newblock \emph{\bibinfo{title}{Non-Equilibrium Phase Transitions}}
  (\bibinfo{publisher}{Springer}, \bibinfo{address}{Dordrecht, The
  Netherlands}, \bibinfo{year}{2008}).

\bibitem{pruessnerSOC2012}
\bibinfo{author}{Pruessner, G.}
\newblock \emph{\bibinfo{title}{Self-Organised Criticality}}
  (\bibinfo{publisher}{Cambridge University Press},
  \bibinfo{address}{Cambridge, UK}, \bibinfo{year}{2012}).

\bibitem{munozExpAbsState1999}
\bibinfo{author}{Mu{\~n}oz, M.~A.}, \bibinfo{author}{Dickman, R.},
  \bibinfo{author}{Vespignani, A.} \& \bibinfo{author}{Zapperi, S.}
\newblock \bibinfo{title}{Avalanche and spreading exponents in systems with
  absorbing states}.
\newblock \emph{\bibinfo{journal}{Phys. Rev. E}}
  \textbf{\bibinfo{volume}{59(5)}}, \bibinfo{pages}{6175--6179}
  \urlprefix\url{10.1103/PhysRevE.59.6175} (\bibinfo{year}{1999}).

\bibitem{odorReview2004}
\bibinfo{author}{{\'O}dor, G.}
\newblock \bibinfo{title}{Universality classes in nonequilibrium lattice
  systems}.
\newblock \emph{\bibinfo{journal}{Rev. Mod. Phys.}}
  \textbf{\bibinfo{volume}{76(3)}}, \bibinfo{pages}{663--724}
  (\bibinfo{year}{2004}).

\bibitem{plenzExp2013}
\bibinfo{author}{Yu, S.}, \bibinfo{author}{Yang, H.}, \bibinfo{author}{Shriki,
  O.} \& \bibinfo{author}{Plenz, D.}
\newblock \bibinfo{title}{Universal organization of resting brain activity at
  the thermodynamic critical point.}
\newblock \emph{\bibinfo{journal}{Front. Syst. Neurosci.}}
  \textbf{\bibinfo{volume}{7}} \urlprefix\url{10.3389/fnsys.2013.00042} (\bibinfo{year}{2013}).

\bibitem{munozGrif2013}
\bibinfo{author}{Moretti, P.} \& \bibinfo{author}{Mu{\~n}oz, M.~A.}
\newblock \bibinfo{title}{Griffiths phases and the stretching of criticality in
  brain networks}.
\newblock \emph{\bibinfo{journal}{Nat. Commun.}} \textbf{\bibinfo{volume}{4}},
  \bibinfo{pages}{2521} \urlprefix\url{10.1038/ncomms3521} (\bibinfo{year}{2013}).

\bibitem{griffiths1969}
\bibinfo{author}{Griffiths, R.~B.}
\newblock \bibinfo{title}{Nonanalytic behavior above the critical point in a
  random {I}sing ferromagnet}.
\newblock \emph{\bibinfo{journal}{Phys. Rev. Lett.}}
  \textbf{\bibinfo{volume}{23(1)}}, \bibinfo{pages}{17--19}
  \urlprefix\url{10.1103/PhysRevLett.23.17} (\bibinfo{year}{1969}).

\bibitem{vojtaGrifRev2006}
\bibinfo{author}{Vojta, T.}
\newblock \bibinfo{title}{Rare region effects at classical, quantum and
  nonequilibrium phase transitions}.
\newblock \emph{\bibinfo{journal}{J. Phys. A: Math. Gen.}}
  \textbf{\bibinfo{volume}{39}}, \bibinfo{pages}{R143} \urlprefix\url{10.1088/0305-4470/39/22/R01} (\bibinfo{year}{2006}).

\bibitem{williamsV12002}
\bibinfo{author}{Williams, S.~R.} \& \bibinfo{author}{Stuart, G.~J.}
\newblock \bibinfo{title}{Dependence of {EPSP} efficacy on synapse location in
  neocortical pyramidal neurons}.
\newblock \emph{\bibinfo{journal}{Science}} \textbf{\bibinfo{volume}{295}},
  \bibinfo{pages}{1907--1910} (\bibinfo{year}{2002}).

\bibitem{songEPSP2005}
\bibinfo{author}{Song, S.}, \bibinfo{author}{Sj{\"o}str{\"o}m, P.~J.},
  \bibinfo{author}{Reigl, M.}, \bibinfo{author}{Nelson, S.} \&
  \bibinfo{author}{Chklovskii, D.~B.}
\newblock \bibinfo{title}{Highly nonrandom features of synaptic connectivity in
  local cortical circuits}.
\newblock \emph{\bibinfo{journal}{PLoS Biol.}} \textbf{\bibinfo{volume}{3(3)}},
  \bibinfo{pages}{e68} \urlprefix\url{10.1371/journal.pbio.0030068} (\bibinfo{year}{2005}).

\bibitem{lefortEPSP2009}
\bibinfo{author}{Lefort, S.}, \bibinfo{author}{Tomm, C.},
  \bibinfo{author}{Floyd{~}Sarria, J.-C.} \& \bibinfo{author}{Petersen, C.~C.}
\newblock \bibinfo{title}{The excitatory neuronal network of the c2 barrel
  column in mouse primary somatosensory cortex}.
\newblock \emph{\bibinfo{journal}{Neuron}} \textbf{\bibinfo{volume}{61(2)}},
  \bibinfo{pages}{301--316} \urlprefix\url{10.1016/j.neuron.2008.12.020} (\bibinfo{year}{2009}).

\bibitem{janssenEPDynP2004}
\bibinfo{author}{Janssen, H.-K.}, \bibinfo{author}{M{\"u}ller, M.} \&
  \bibinfo{author}{Stenull, O.}
\newblock \bibinfo{title}{Generalized epidemic process and tricritical dynamic
  percolation}.
\newblock \emph{\bibinfo{journal}{Phys. Rev. E}} \textbf{\bibinfo{volume}{70}},
  \bibinfo{pages}{026114} \urlprefix\url{10.1103/PhysRevE.70.026114} (\bibinfo{year}{2004}).

\bibitem{novikov1fB1997}
\bibinfo{author}{Novikov, E.}, \bibinfo{author}{Novikov, A.},
  \bibinfo{author}{Shannahoff-{K}halsa, D.}, \bibinfo{author}{Schwartz, B.} \&
  \bibinfo{author}{Wright, J.}
\newblock \bibinfo{title}{Scale-similar activity in the brain}.
\newblock \emph{\bibinfo{journal}{Phys. Rev. E}}
  \textbf{\bibinfo{volume}{56(3)}}, \bibinfo{pages}{R2387--R2389}
  (\bibinfo{year}{1997}).

\bibitem{teichcat97}
\bibinfo{author}{Teich, M.~C.}, \bibinfo{author}{Heneghan, C.},
  \bibinfo{author}{Lowen, S.~B.}, \bibinfo{author}{Ozaki, T.} \&
  \bibinfo{author}{Kaplan, E.}
\newblock \bibinfo{title}{Fractal character of the neural spike train in the
  visual system of the cat}.
\newblock \emph{\bibinfo{journal}{J. Opt. Soc. Am. A}}
  \textbf{\bibinfo{volume}{14(3)}}, \bibinfo{pages}{529--546}
  \urlprefix\url{10.1364/JOSAA.14.000529} (\bibinfo{year}{1997}).

\bibitem{henrie1fV12005}
\bibinfo{author}{Andrew{~}{H}enrie, J.} \& \bibinfo{author}{Shapley, R.}
\newblock \bibinfo{title}{{LFP} power spectra in {V}1 cortex: The graded effect
  of stimulus contrast}.
\newblock \emph{\bibinfo{journal}{J. Neurophysiol.}}
  \textbf{\bibinfo{volume}{94}}, \bibinfo{pages}{479--490}
  \urlprefix\url{10.1152/jn.00919.2004} (\bibinfo{year}{2005}).

\bibitem{hermes1fV12014}
\bibinfo{author}{Hermes, D.}, \bibinfo{author}{Miller, K.~J.},
  \bibinfo{author}{Wandell, B.~A.} \& \bibinfo{author}{Winawer, J.}
\newblock \bibinfo{title}{Stimulus dependence of gamma oscillations in human
  visual cortex}.
\newblock \emph{\bibinfo{journal}{Cereb. Cortex}}  \urlprefix\url{10.1093/cercor/bhu091} (\bibinfo{year}{2014}).

\bibitem{stanleyDFA1995}
\bibinfo{author}{Peng, C.-K.}, \bibinfo{author}{Havlin, S.},
  \bibinfo{author}{Stanley, H.~E.} \& \bibinfo{author}{Goldberger, A.~L.}
\newblock \bibinfo{title}{Quantification of scaling exponents and crossover
  phenomena in nonstationary heartbeat time series}.
\newblock \emph{\bibinfo{journal}{Chaos}} \textbf{\bibinfo{volume}{5}},
  \bibinfo{pages}{82--87} \urlprefix\url{http://dx.doi.org/10.1063/1.166141} (\bibinfo{year}{1995}).

\bibitem{linkenkaerDFA2012}
\bibinfo{author}{Hardstone, R.} \emph{et~al.}
\newblock \bibinfo{title}{Detrended fluctuation analysis: a scale-free view on
  neuronal oscillations}.
\newblock \emph{\bibinfo{journal}{Front. Physiol.}}
  \textbf{\bibinfo{volume}{3}}, \bibinfo{pages}{450} \urlprefix\url{10.3389/fphys.2012.00450} (\bibinfo{year}{2012}).

\bibitem{hartleyISuscep2013}
\bibinfo{author}{Taylor, T.~J.}, \bibinfo{author}{Hartley, C.},
  \bibinfo{author}{Simon, P.~L.}, \bibinfo{author}{Kiss, I.~Z.} \&
  \bibinfo{author}{Berthouze, L.}
\newblock \bibinfo{title}{Identification of criticality in neuronal avalanches:
  {I}. {A} theoretical investigation of the non-driven case}.
\newblock \emph{\bibinfo{journal}{J. Math. Neurosci.}}
  \textbf{\bibinfo{volume}{3}}, \bibinfo{pages}{5} \urlprefix\url{10.1186/2190-8567-3-5} (\bibinfo{year}{2013}).

\bibitem{destexhePL2015}
\bibinfo{author}{Touboul, J.} \& \bibinfo{author}{Destexhe, A.}
\newblock \bibinfo{title}{Power-law statistics and universal scaling in the
  absence of criticality}.
\newblock \emph{\bibinfo{journal}{arXiv:1503.08033 [q-bio.NC]}}
  (\bibinfo{year}{2015}).

\bibitem{okuskyV11982}
\bibinfo{author}{O{'}{K}usky, J.} \& \bibinfo{author}{Colonnier, M.}
\newblock \bibinfo{title}{A laminar analysis of the number of neurons, glia and
  synapses in the visual cortex (area 17) of adult macaque monkeys}.
\newblock \emph{\bibinfo{journal}{J. Comp. Neurol.}}
  \textbf{\bibinfo{volume}{210}}, \bibinfo{pages}{278--290}
  (\bibinfo{year}{1982}).

\bibitem{lundV11984}
\bibinfo{author}{Lund, J.~S.}
\newblock \bibinfo{title}{Spiny stellate neurons}.
\newblock In \bibinfo{editor}{Peters, A.} \& \bibinfo{editor}{Jones, E.~G.}
  (eds.) \emph{\bibinfo{booktitle}{The Cerebral Cortex}},
  vol.~\bibinfo{volume}{1}, chap.~\bibinfo{chapter}{7},
  \bibinfo{pages}{255--308} (\bibinfo{publisher}{Plenum Press},
  \bibinfo{address}{New York, USA}, \bibinfo{year}{1984}).

\bibitem{callawayV1Review1998}
\bibinfo{author}{Callaway, E.~M.}
\newblock \bibinfo{title}{Local circuits in primary visual cortex of the
  macaque monkey}.
\newblock \emph{\bibinfo{journal}{Annu. Rev. Neurosci.}}
  \textbf{\bibinfo{volume}{21(2)}}, \bibinfo{pages}{47--74}
  (\bibinfo{year}{1998}).

\bibitem{yabutaV11998}
\bibinfo{author}{Yabuta, N.~H.} \& \bibinfo{author}{Callaway, E.~M.}
\newblock \bibinfo{title}{Functional streams and local connections of layer 4c
  neurons in primary visual cortex of the macaque monkey}.
\newblock \emph{\bibinfo{journal}{J. Neurosci.}}
  \textbf{\bibinfo{volume}{18(22)}}, \bibinfo{pages}{9489--9499}
  (\bibinfo{year}{1998}).

\bibitem{albrightNeuroRev2000}
\bibinfo{author}{Albright, T.~D.}, \bibinfo{author}{Jessell, T.~M.},
  \bibinfo{author}{Kandel, E.~R.} \& \bibinfo{author}{Posner, M.~I.}
\newblock \bibinfo{title}{Neural science: A century of progress and the
  mysteries that remain}.
\newblock \emph{\bibinfo{journal}{Neuron}} \textbf{\bibinfo{volume}{25}},
  \bibinfo{pages}{S1--S55} \urlprefix\url{10.1016/S0896-6273(00)80912-5} (\bibinfo{year}{2000}).

\bibitem{levina}
\bibinfo{author}{Levina, A.}, \bibinfo{author}{Herrmann, J.~M.} \&
  \bibinfo{author}{Geisel, T.}
\newblock \bibinfo{title}{Dynamical synapses causing self-organized criticality
  in neural networks}.
\newblock \emph{\bibinfo{journal}{Nat. Phys.}} \textbf{\bibinfo{volume}{3}},
  \bibinfo{pages}{857--860} (\bibinfo{year}{2007}).

\bibitem{ariadneDynSyn2015}
\bibinfo{author}{de~Andrade{ }Costa, A.}, \bibinfo{author}{Copelli, M.} \&
  \bibinfo{author}{Kinouchi, O.}
\newblock \bibinfo{title}{Can dynamical synapses produce true self-organized
  criticality?}
\newblock \emph{\bibinfo{journal}{J. Stat. Mech.}} \bibinfo{pages}{P06004}
  \urlprefix\url{10.1088/1742-5468/2015/06/P06004} (\bibinfo{year}{2015}).

\bibitem{girardiAva}
\bibinfo{author}{Girardi-Schappo, M.}, \bibinfo{author}{Kinouchi, O.} \&
  \bibinfo{author}{Tragtenberg, M. H.~R.}
\newblock \bibinfo{title}{Critical avalanches and subsampling in map-based
  neural networks coupled with noisy synapses}.
\newblock \emph{\bibinfo{journal}{Phys. Rev. E}} \textbf{\bibinfo{volume}{88}},
  \bibinfo{pages}{024701} \urlprefix\url{10.1103/PhysRevE.88.024701} (\bibinfo{year}{2013}).

\bibitem{beggsSOqC2014}
\bibinfo{author}{Williams-Garc\'{i}a, R.~V.}, \bibinfo{author}{Moore, M.},
  \bibinfo{author}{Beggs, J.~M.} \& \bibinfo{author}{Ortiz, G.}
\newblock \bibinfo{title}{Quasicritical brain dynamics on a nonequilibrium
  widom line}.
\newblock \emph{\bibinfo{journal}{Phys. Rev. E}}
  \textbf{\bibinfo{volume}{90(6)}}, \bibinfo{pages}{062714}
  \urlprefix\url{10.1103/PhysRevE.90.062714} (\bibinfo{year}{2014}).

\bibitem{andreazzaSim2006}
\bibinfo{author}{Andreazza, J.~K.} \& \bibinfo{author}{Pinto, L.~T.}
\newblock \bibinfo{title}{Simulation of the primary visual cortex of the
  macaque monkey by natural neural networks}.
\newblock In \emph{\bibinfo{booktitle}{Prooceedings of 2nd LNCC Meeting on
  Computational Modelling}} (\bibinfo{address}{Petr{\'o}polis, RJ, Brazil},
  \bibinfo{year}{2006}).

\bibitem{kandel2012}
\bibinfo{author}{Kandel, E.}, \bibinfo{author}{Schwartz, J.} \&
  \bibinfo{author}{Jessell, T.}
\newblock \emph{\bibinfo{title}{Principles of Neural Science}}
  (\bibinfo{publisher}{McGraw-Hill Education}, \bibinfo{address}{Columbus, OH,
  USA}, \bibinfo{year}{2012}).

\bibitem{deutschNervous1993}
\bibinfo{author}{Deutsch, S.} \& \bibinfo{author}{Deutsch, A.}
\newblock \emph{\bibinfo{title}{Understanding the Nervous System}}
  (\bibinfo{publisher}{{I}{E}{E}{E} Press}, \bibinfo{address}{New York, NY,
  USA}, \bibinfo{year}{1993}).

\bibitem{girardiV1conf2015}
\bibinfo{author}{Bortolotto, G.~S.}, \bibinfo{author}{Girardi{-}Schappo, M.},
  \bibinfo{author}{Gonsalves, J.~J.}, \bibinfo{author}{Pinto, L.~T.} \&
  \bibinfo{author}{Tragtenberg, M. H.~R.}
\newblock \bibinfo{title}{Information processing occurs via critical avalanches
  in a model of the primary visual cortex}.
\newblock \emph{\bibinfo{journal}{J. Phys. Conf. Ser.}}
  \textbf{\bibinfo{volume}{686}}, \bibinfo{pages}{012008}
  \urlprefix\url{10.1088/1742-6596/686/1/012008} (\bibinfo{year}{2016}).

\bibitem{grassbergerGEP1983}
\bibinfo{author}{Grassberger, P.}
\newblock \bibinfo{title}{On the critical behavior of the general epidemic
  process and dynamical percolation}.
\newblock \emph{\bibinfo{journal}{Math. Biosci.}}
  \textbf{\bibinfo{volume}{63(2)}}, \bibinfo{pages}{157--172}
  \urlprefix\url{10.1016/0025-5564(82)90036-0} (\bibinfo{year}{1983}).

\bibitem{ferreiraSus2012}
\bibinfo{author}{Ferreira, S.~C.}, \bibinfo{author}{Castellano, C.} \&
  \bibinfo{author}{Pastor-Satorras, R.}
\newblock \bibinfo{title}{Epidemic thresholds of the
  susceptible-infected-susceptible model on networks: A comparison of numerical
  and theoretical results}.
\newblock \emph{\bibinfo{journal}{Phys. Rev. E}} \textbf{\bibinfo{volume}{86}},
  \bibinfo{pages}{041125} \urlprefix\url{10.1103/PhysRevE.86.041125} (\bibinfo{year}{2012}).

\bibitem{teramaeFukai2007}
\bibinfo{author}{Teramae, J.-N.} \& \bibinfo{author}{Fukai, T.}
\newblock \bibinfo{title}{Local cortical circuit model inferred from power-law
  distributed neuronal avalanches}.
\newblock \emph{\bibinfo{journal}{J. Comput. Neurosci.}}
  \textbf{\bibinfo{volume}{22(3)}}, \bibinfo{pages}{301--312}
  \urlprefix\url{10.1007/s10827-006-0014-6} (\bibinfo{year}{2007}).

\bibitem{brayGrif1988}
\bibinfo{author}{Bray, A.~J.} \& \bibinfo{author}{Rodgers, G.~J.}
\newblock \bibinfo{title}{Diffusion in a sparsely connected space: A model for
  glassy relaxation}.
\newblock \emph{\bibinfo{journal}{Phys. Rev. B}}
  \textbf{\bibinfo{volume}{38(16)}}, \bibinfo{pages}{11461--11470}
  (\bibinfo{year}{1988}).

\bibitem{chungRew2002}
\bibinfo{author}{Chung, S.}, \bibinfo{author}{Li, X.} \&
  \bibinfo{author}{Nelson, S.~B.}
\newblock \bibinfo{title}{Short-term depression at thalamocortical synapses
  contributes to rapid adaptation of cortical sensory responses in vivo}.
\newblock \emph{\bibinfo{journal}{Neuron}} \textbf{\bibinfo{volume}{34}},
  \bibinfo{pages}{437--446} \urlprefix\url{10.1016/S0896-6273(02)00659-1} (\bibinfo{year}{2002}).

\bibitem{alamancosRew2004}
\bibinfo{author}{Castro{-}Alamancos, M.~A.}
\newblock \bibinfo{title}{Absence of rapid sensory adaptation in neocortex
  during information processing states}.
\newblock \emph{\bibinfo{journal}{Neuron}} \textbf{\bibinfo{volume}{41}},
  \bibinfo{pages}{455--464} \urlprefix\url{10.1016/S0896-6273(03)00853-5}
\newblock  (\bibinfo{year}{2004}).

\end{thebibliography}
 \newcommand{\noop}[1]{}

\section*{Acknowledgements}

The model and the first version of its code have been conceived and implemented by J.~K.~Andreazza and LTP.
We thank J.-P.~Thivierge for providing the code for Maximum Likelihood test.
We also thank M.~Copelli, O.~Kinouchi, R.~Dickman, S.~Boettcher and T.~B.~Pedro for discussions.
MGS and GSB thank the partial financial support of CNPq and FAPESC, respectively.

\section*{Author contributions statement}

LTP conceived the model;
MGS and LTP updated the code;
MGS, GSB and LTP ran the simulations;
All authors analyzed the results;
MGS and GSB wrote the manuscript;
All authors reviewed the manuscript.

\section*{Additional information}

\textbf{Supplementary information} accompanies this paper at http://www.nature.com/srep

\textbf{Competing financial interests}: The authors declare no competing financial interests.

\end{document}